%% file: ms.tex
\documentclass[usenatbib]{mn2e}
\usepackage{natbibmnfix,graphicx,times, amsmath}

\include{defns}

\def\myputfigure#1#2#3#4#5%
{\vskip#5pt\makebox[0pt]{\hskip#2in
\includegraphics[width=#3\textwidth]{#1}}\vskip#4pt\hfill}

\newcommand\lsim{\mathrel{\rlap{\lower4pt\hbox{\hskip1pt$\sim$}}
        \raise1pt\hbox{$<$}}}
\newcommand\gsim{\mathrel{\rlap{\lower4pt\hbox{\hskip1pt$\sim$}}
        \raise1pt\hbox{$>$}}}

\begin{document}


\title[21cmFAST]{21cmFAST: A Fast, Semi-Numerical Simulation of the High-Redshift 21-cm Signal}

\author[Mesinger et al.]{Andrei Mesinger$^1$\thanks{Hubble Fellow; email: mesinger@astro.princeton.edu}, Steven Furlanetto$^2$, \& Renyue Cen$^1$ \\
$^1$Department of Astrophysical Sciences, Princeton University, Princeton, NJ 08544, USA \\
$^2$Department of Physics and Astronomy, University of California, Los Angeles, CA 90095, USA}

\voffset-.6in

\maketitle

\begin{abstract}
We introduce a powerful semi-numeric modeling tool, 21cmFAST, designed to efficiently simulate the cosmological 21-cm signal. Our code generates 3D realizations of evolved density, ionization, peculiar velocity, and spin temperature fields, which it then combines to compute the 21-cm brightness temperature.
Although the physical processes are treated with approximate methods, we compare our results to a state-of-the-art large-scale hydrodynamic simulation, and find good agreement on scales pertinent to the upcoming observations ($\gsim$ 1 Mpc).  The power spectra from 21cmFAST agree with those generated from the numerical simulation to within 10s of percent, down to the Nyquist frequency.
  We show results from a 1 Gpc simulation which tracks the cosmic 21-cm signal down from $z=250$, highlighting the various interesting epochs.  Depending on the desired resolution, 21cmFAST can compute a redshift realization on a single processor in just a few minutes.
  Our code is fast, efficient, customizable and publicly available, making it a useful tool for 21-cm parameter studies.
\end{abstract}


\section{Introduction}
\label{sec:intro}

Through challenging observational efforts, the high-redshift frontier has been incrementally pushed back in recent years.  Glimpses of the $z\sim$ 6--8 Universe were provided by quasars (e.g. \citealt{Fan06}), candidate Lyman break galaxies (e.g. \citealt{Bouwens08, McLure09, Bouwens09, Ouchi09}), Lyman alpha emitters (LAEs; e.g. \citealt{Shimasaku06, Kashikawa06}), and GRBs (e.g. \citealt{Cusumano06, Greiner09, Salvaterra09}).  Unfortunately, these precious observations currently provide only a 
limited set of relatively bright objects.
Luckily, we will soon be inundated with observations probing this and even earlier epochs.
 These observations should include infrared spectra from the {\it James Webb Space Telescope} (JWST), the Thirty Meter Telescope (TMT), the Giant Magellan Telescope (GMT), the European Extremely Large Telescope (E-ELT), wide-field LAE surveys from the Subaru HyperSupremeCam, as well as the E-mode CMB polarization power spectrum measured by the {\it Planck} satellite.  Some of the most important information will come in  the form of the redshifted 21-cm line of neutral hydrogen.  Several interferometers will attempt to observe the cosmological 21-cm signal, including the Mileura Wide Field Array (MWA; \citealt{Bowman05})\footnote{http://web.haystack.mit.edu/arrays/MWA/}, the Low Frequency Array (LOFAR)\footnote{http://www.lofar.org}, the Giant Metrewave Radio Telescope (GMRT; \citealt{Pen08}), the Precision Array to Probe the Epoch of Reionization (PAPER; \citealt{Parsons09}), and eventually the Square Kilometer Array (SKA)\footnote{http://www.skatelescope.org/}.
  The first generation of these instruments, most notably LOFAR and MWA, are not only scheduled to become operational within a year, but should also yield insight into the 3D distribution of HI, provided the systematics can be overcome (see the recent reviews of \citealt{FOB06, MW09}).

However, interpreting this data will be quite challenging and no-doubt controversial, as foreshadowed by the confusion surrounding the scant, currently-available observations. There are two main challenges to overcome: (1) an extremely large parameter space, due to our poor understanding of the high-redshift Universe; (2) an enormous dynamic range (i.e. range of relevant scales).

Theoretically, the dawn of the first astrophysical objects and reionization could be modeled from first principles using numerical simulations, which include the complex interplay of many physical processes.  In practice however, simulating these epochs requires enormous simulation boxes.   Gigaparsec scales are necessary to statistically model ionized regions and absorption systems or create accurate mock spectra from the very rare high-redshift quasars.  However, the simulations also require high enough resolution to resolve the underlying sources and sinks of ionizing photons and the complex small-scale feedback mechanisms which regulate them.  Thus one is forced to make compromises: deciding which physical processes can be ignored, and how the others can be parameterized and efficiently folded into large-scale models. Furthermore, large-scale 
simulations are computationally costly (even when they sacrifice completeness for speed by ignoring hydrodynamic processes) and thus are inefficient in large parameter studies.

On the other hand, analytic models, while more approximate, are fast and can provide physical insight into the import of various processes.  However, analytical models are hard-pressed to go beyond the linear regime, and 
beyond making fairly simple predictions such as the mean 21-cm signal \citep{Furlanetto06}, the probability density function (PDF; \citealt{FZH04_21cmtop}) and power spectrum \citep{PF07, Barkana09}.  The 21-cm tomographic signal should be rich in information, accommodating many additional, higher-order statistical probes, such as the bi-spectrum (Pritchard et al., in preparation), the difference PDF \citep{BL08}, etc.

In this paper, we follow a path of compromise, attempting to preserve the most useful elements of both analytic and numeric approaches.  We introduce a self-consistent, semi-numerical\footnote{By ``semi-numerical'' we mean using more approximate physics than numerical simulations, but capable of independently generating 3D realizations.} simulation, specifically optimized to predict the high-redshift 21-cm signal.  Through a combination of the excursion-set formalism and perturbation theory, our code can generate full 3D realizations of the density, ionization, velocity, spin temperature, and ultimately 21-cm brightness temperature fields.  Although the physical processes are treated with approximate methods, our results agree well with a state-of-the-art hydrodynamic simulation of reionization.
  However, unlike numerical simulations, realizations are computationally cheap and can be generated in a matter of minutes on a single processor, with modest memory requirements.  Most importantly, our code is publicly available at http://www.astro.princeton.edu/$\sim$mesinger/Sim.html.  We name our simulation {\it 21cmFAST}.

Semi-Numerical approaches have already proved invaluable in reionization studies \citep{Zahn05, MF07, GW08, Alvarez09, CHR09, Thomas09}.  Indeed, 21cmFAST is a more specialized version of our previous code, DexM (\citealt{MF07}; hereafter MF07).  The difference between the two is that 21cmFAST bypasses the halo finding algorithm, resulting in a faster code with a larger dynamic range and more modest memory requirements.  In this work, we also introduce some new additions to our code, mainly to compute the spin temperature.

In \S \ref{sec:post_heat}, we compare predictions from 21cmFAST with those from hydrodynamic simulations of the various physical components comprising the 21-cm signal in the post heating regime.  Density, ionization, peculiar velocity gradient, and full 21-cm brightness temperature fields are explored in \S \ref{sec:den}, \S \ref{sec:ion}, \S \ref{sec:dvdr}, \S \ref{sec:21cm_cmp}, respectively.  In \S \ref{sec:heating}, we introduce our method for computing the spin temperature fields, with results from the complete calculation (including the spin temperature) presented in \S \ref{sec:results}.  Finally in \S \ref{sec:conc}, we summarize our findings.

Unless stated otherwise, we quote all quantities in comoving units. We adopt the background cosmological parameters ($\Omega_\Lambda$, $\Omega_{\rm M}$, $\Omega_b$, $n$, $\sigma_8$, $H_0$) = (0.72, 0.28, 0.046, 0.96, 0.82, 70 km s$^{-1}$ Mpc$^{-1}$), matching the five--year results of the {\it WMAP} satellite \citep{Komatsu09}.

\section{21-cm Temperature Fluctuations Post Heating ($T_S \gg \Tcmb$)}
\label{sec:post_heat}

Our ultimate goal is to compute the 21 cm background, which requires a number of physics components.  To identify them, note that the offset of the 21-cm brightness temperature from the CMB temperature, $\Tcmb$, along a line of sight (LOS) at observed frequency $\nu$, can be written as (c.f. \citealt{FOB06}):

\begin{align}
\label{eq:delT}
\nonumber \delT(\nu) = &\frac{\Ts - \Tcmb}{1+z} (1 - e^{-\tau_{\nu_0}}) \approx \\
\nonumber &27 \nf (1+\delNL) \left(\frac{H}{dv_r/dr + H}\right) \left(1 - \frac{\Tcmb}{\Ts} \right) \\
&\times \left( \frac{1+z}{10} \frac{0.15}{\Omega_{\rm M} h^2}\right)^{1/2} \left( \frac{\Omega_b h^2}{0.023} \right) {\rm mK},
\end{align}
 
\noindent where $T_S$ is the gas spin temperature, $\tau_{\nu_0}$ is the optical depth at the 21-cm frequency $\nu_0$, $\delNL({\bf x}, z) \equiv \rho/\bar{\rho} - 1$ is the evolved (Eulerian) density contrast, $H(z)$ is the Hubble parameter, $dv_r/dr$ is the comoving gradient of the line of sight component of the comoving velocity, and all quantities are evaluated at redshift $z=\nu_0/\nu - 1$.  The final approximation makes the assumption that that $dv_r/dr \ll H$, which is generally true for the pertinent redshifts and scales, though we shall return to this issue in \S \ref{sec:dvdr}.  

{\it In this comparison section, we make the standard, simplifying assumption of working in the post-heating regime: $T_S \gg \Tcmb$.}   For fiducial astrophysical models, this is likely a safe assumption during the bulk of reionization \citep{Furlanetto06, CM08, Santos08, Baek09}.  We will however revisit this assumption in \S \ref{sec:heating}, where we introduce our method for computing the spin temperature field.  

The remaining components of eq. \ref{eq:delT} are the density, $\delNL$, the ionization, $\nf$, and the velocity gradient, $dv_r/dr$.  Below, we study these in turn, comparing 21cmFAST to the hydrodynamic cosmological simulation of \citet{TCL08}, {\it using the same initial conditions (ICs)}.  We perform ``by-eye'' comparisons at various redshifts/stages of reionization, as well as one and two-point statistics: the PDFs (smoothed on several scales), and the power spectra.
 Since our code is designed to simulate the cosmological 21-cm signal from neutral hydrogen, we study the regime before the likely completion of reionization, $z\gsim7$ (though present data is even consistent with reionization completing at $z\lsim$6; \citealt{Lidz07, Mesinger09}).

 The simulations of \citet{TCL08} are the current ``state-of-the-art'' reionization simulations. They include simultaneous treatment of dark matter (DM) and gas, five-frequency radiative transfer (RT) on a 512$^3$ grid, and they resolve $M_{\rm halo}\gsim10^8 \Msun$ ionizing sources with $\gsim 40$ DM particles in a 143 Mpc box.

\begin{figure*}
{
\includegraphics[width=0.45\textwidth]{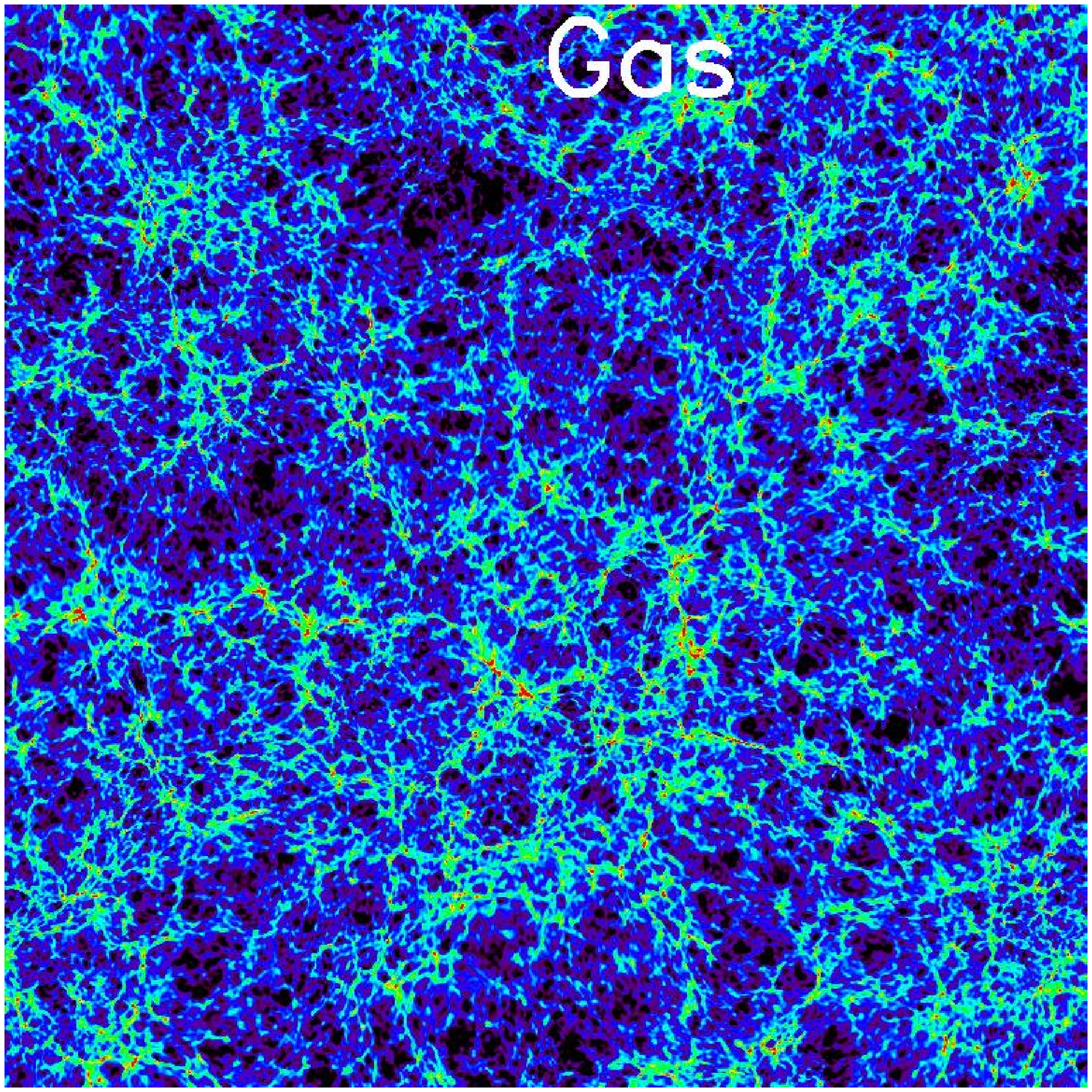}
\includegraphics[width=0.45\textwidth]{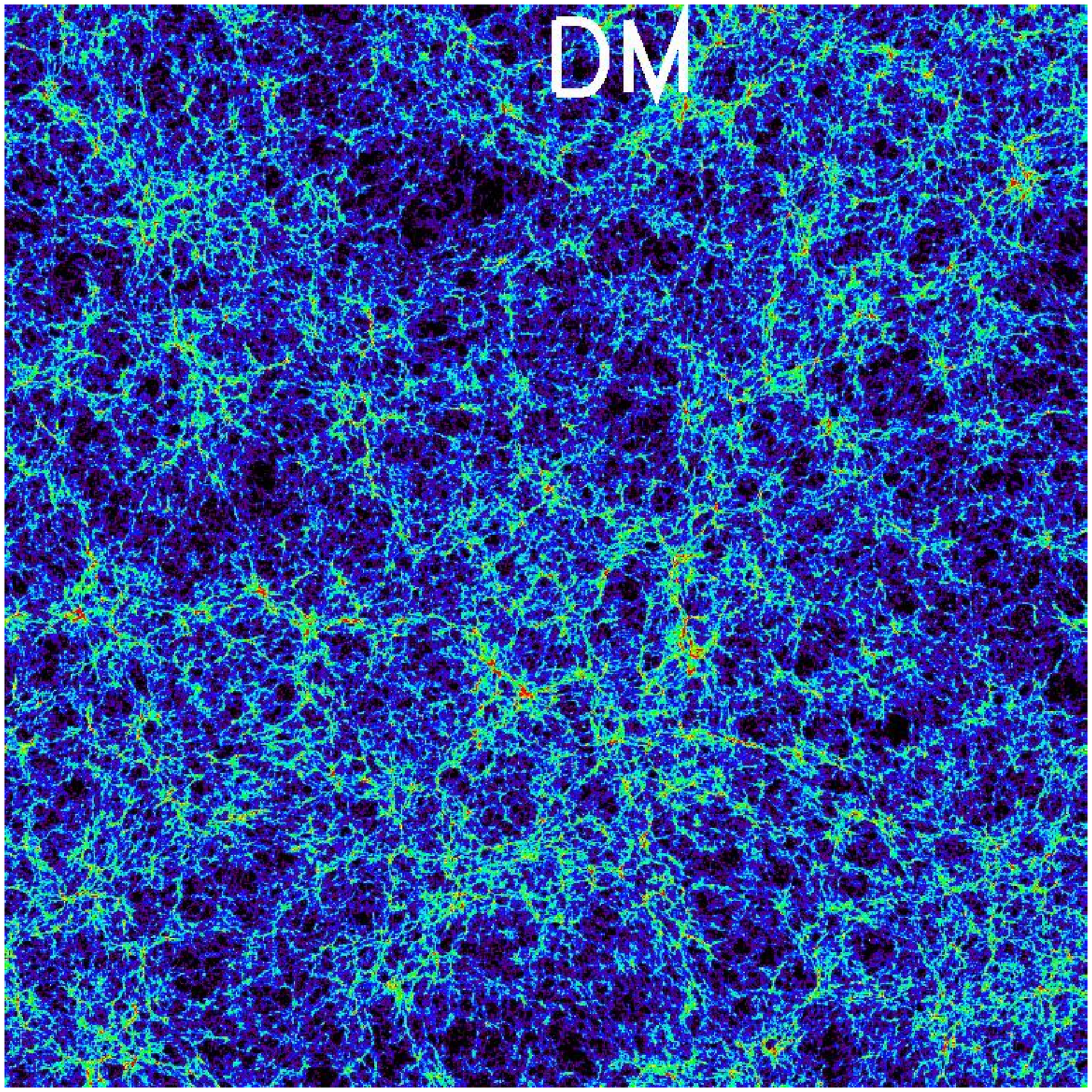}
\includegraphics[width=0.45\textwidth]{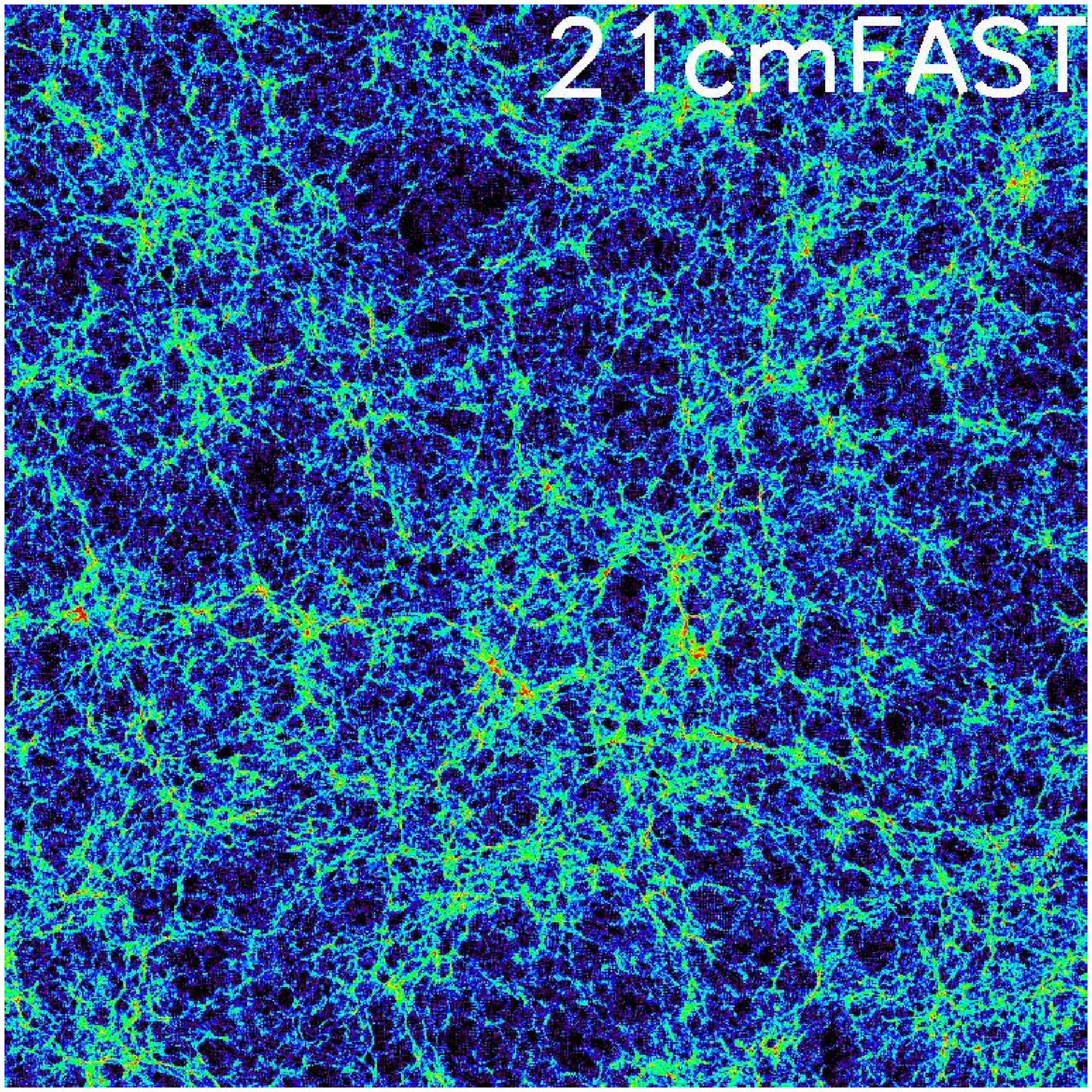}
\includegraphics[width=0.45\textwidth]{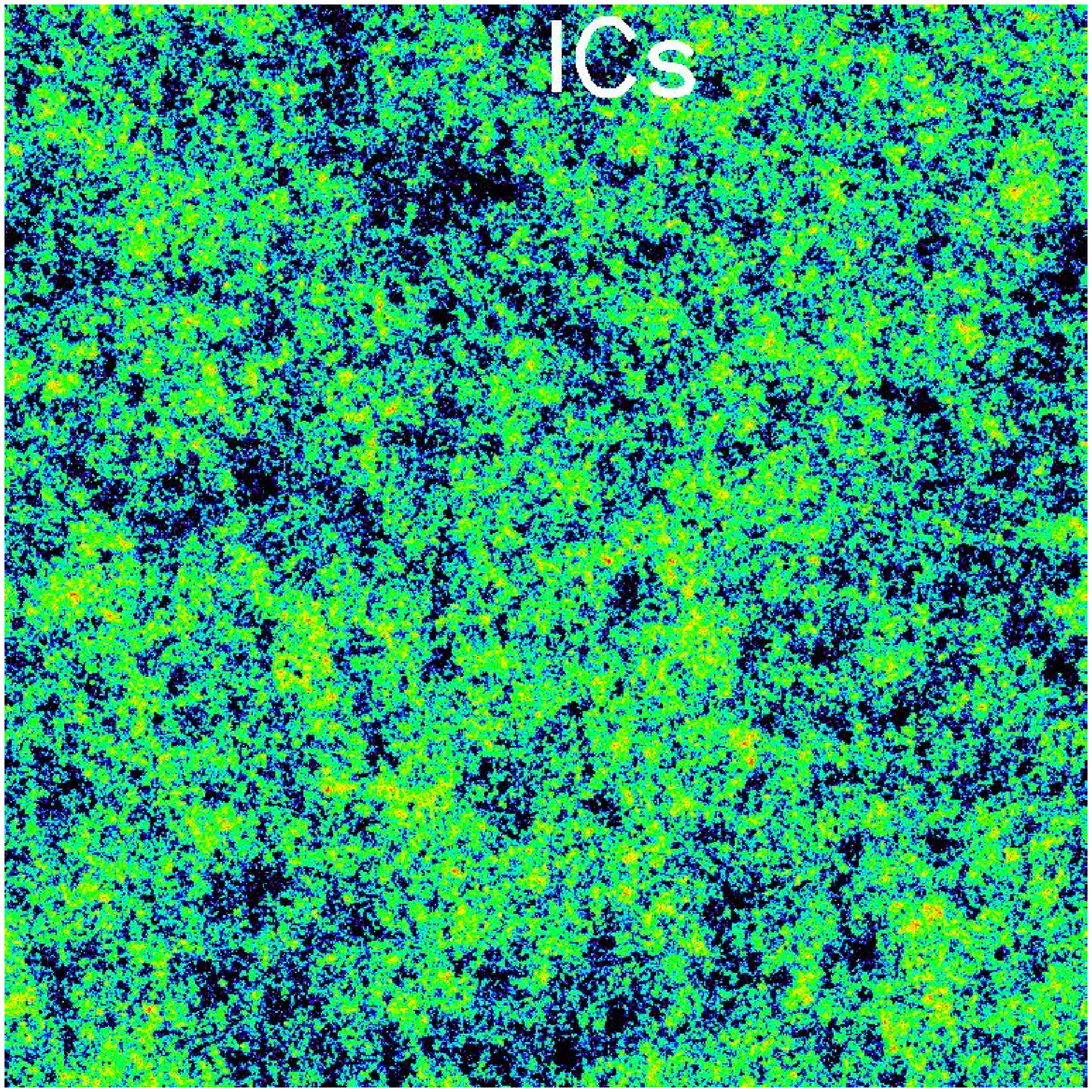}
\vskip0.0pt
}
\includegraphics[width=0.6\textwidth]{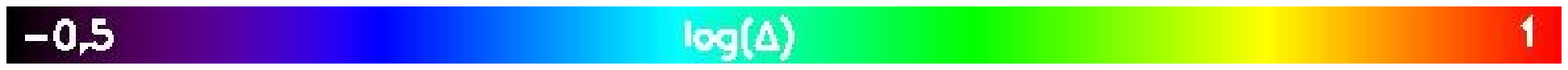}
\caption{
Slices through the density field, $\Delta({\bf x}, z)$, for the gas, DM, linearly-extrapolated ICs, and 21cmFAST, at $z=$ 7, ({\it clockwise from top left}).  Each slice is 143 Mpc on a side and 0.19 Mpc thick.
\label{fig:den_pics}
}
\vspace{-1\baselineskip}
\end{figure*}

\subsection{Evolved Density Field}
\label{sec:den}

\begin{figure*}
\vspace{+0\baselineskip}
{
\includegraphics[width=0.45\textwidth]{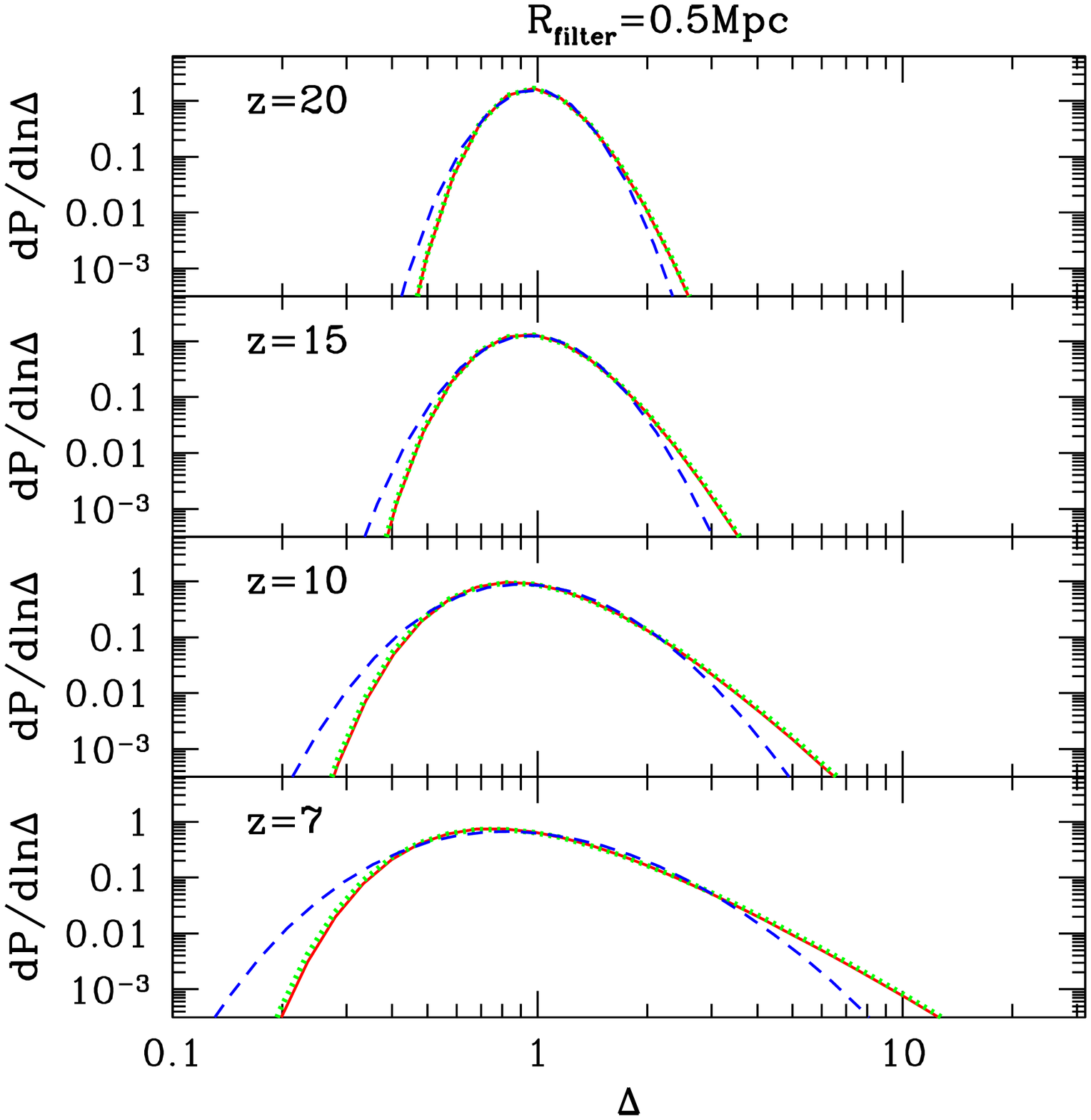}
\includegraphics[width=0.45\textwidth]{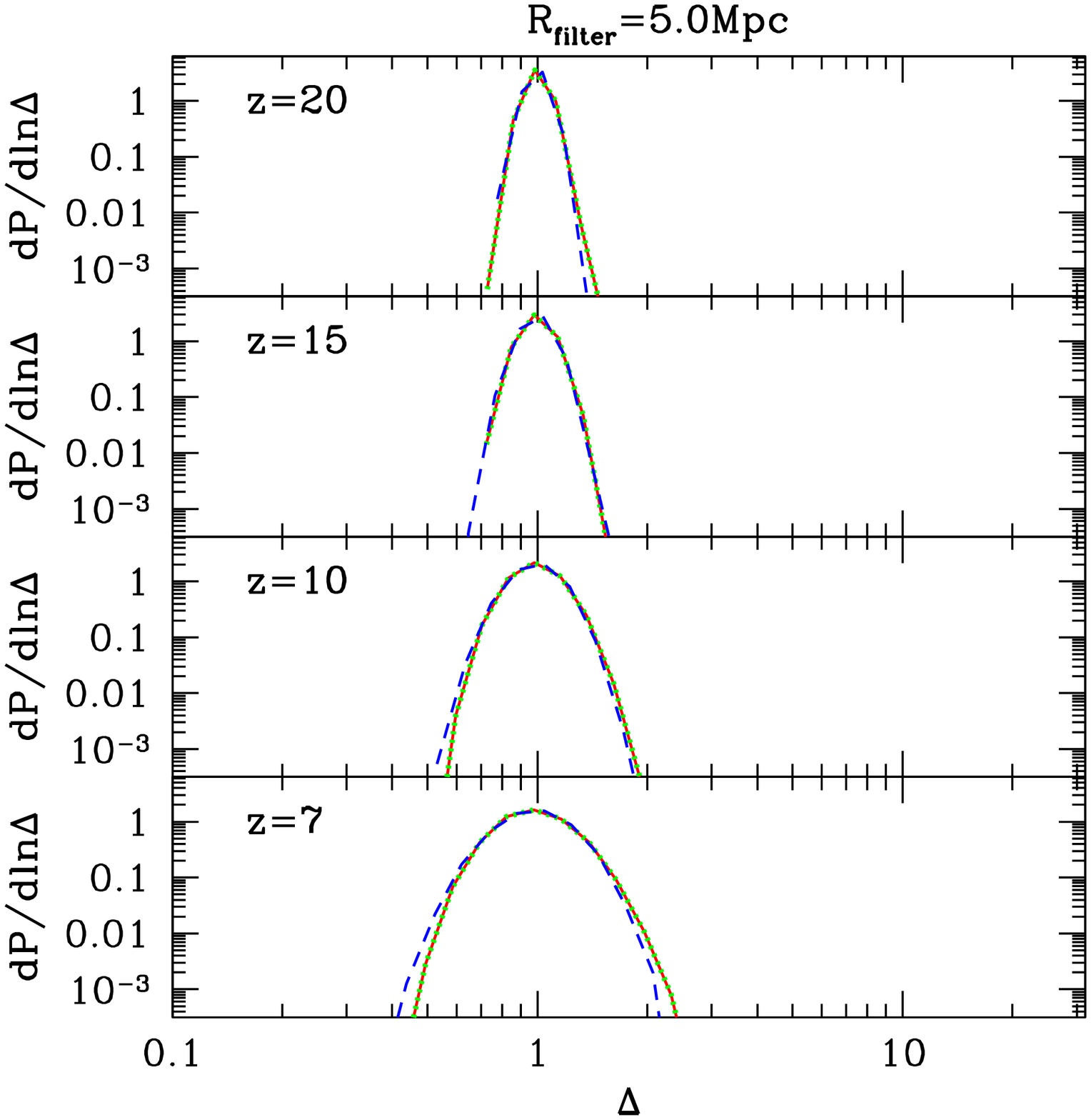}
}
\caption{
PDFs of the density fields smoothed on scale $R_{\rm filter}$, computed from the gas ({\it solid red curves}), DM ({\it dotted green curves}), and 21cmFAST ({\it dashed blue curves}) fields, at $z=$ 20, 15, 10, 7 ({\it top to bottom}).  The left panel corresponds to $R_{\rm filter} = 0.5$ Mpc; the right panel corresponds to $R_{\rm filter}$ = 5 Mpc.  All smoothing was performed with a real-space, top-hat filter.
\label{fig:filter_den_pdfs}
}
\vspace{-1\baselineskip}
\end{figure*}

  We calculate the evolved density field in the same fashion as in the ``parent'' code, DexM, outlined in MF07.  In short, we generate density and velocity ICs in initial (Lagrangian) space, in roughly the same manner as numerical cosmological simulations.  We then approximate gravitational collapse by moving each initial matter particle (whose mass is the total mass in the corresponding IC cell) according to first-order perturbation theory \citep{ZelDovich70}.  First-order perturbation theory is very computationally convenient as the displacement field is a separable function of space and time, so the spatial component need only be computed once for each realization/box.  There is no separate treatment of baryons and DM.
Readers interested in more details concerning this approach are encouraged to check MF07.  

This approach to generating large-scale density fields was also adopted by \citet{CHR09} and \citet{Santos09}, who briefly showed that the resulting fields at high-$z$ traced the DM distribution from an N-body code fairly well.  Here we perform more extensive comparisons.
  The Lagrangian space ICs used for 21cmFAST were initialized at $z=300$ on a 1536$^3$ grid.  The velocity fields used to perturb the ICs, as well as the resulting density fields presented below are 768$^3$.\footnote{Our code allows the ICs to be sampled onto a high-resolution, HI-RES-DIM$^3$, grid, while the subsequent evolved density, ionization, etc. fields can be created at lower resolution, LOW-RES-DIM$^3$.  This allows efficient use of available memory, with the code storing at most HI-RES-DIM$^3$ + 4 $\times$ LOW-RES-DIM$^3$ floating point numbers in RAM.  However, the Zel'Dovich perturbation approach, just as numerical simulations, requires that the evolved fields are adequately resolved by the high-resolution grid.  Failure to do so can substantially underestimate the fluctuations in the density field.  We roughly find that the high-resolution grid should have cell sizes $\lsim 1$ Mpc to accurately model density fields at redshifts of interest ($z\lsim40$).  For larger cell sizes, $\gsim 10$ Mpc, the linear theory evolution option of 21cmFAST should be used.}  We show results from both 768$^3$ and 256$^3$ boxes.
We highlight here that it takes $\sim10$ minutes to generate the $768^3$ 21cmFAST density field from the 1536$^3$ ICs at $z=7$ on a single processor Mac Pro desktop computer.  To put this into perspective, a hydrodynamical simulation of this single realization would take approximately three days to run down to $z=7$ on a 1536-node supercomputing cluster.

In Figure \ref{fig:den_pics}, we show a slice through the evolved density field, $\Delta({\bf x}, z)$ $\equiv$ $1 + \delNL$, at $z=7$.  Density fields computed from the gas, DM, linearly-extrapolated ICs, and 21cmFAST are shown clockwise from the top left panel.
  It is evident that the Zel'Dovich approximation works quite well for this purpose, accurately reproducing the cosmic web.  We do not capture baryonic physics, and so the 21cmFAST output looks more similar to the DM than the gas.  However, hydrodynamics is not included even in most of the present-day cosmological ($\gsim100$ Mpc) simulations (e.g. \citealt{Iliev06_sim, McQuinn07}; see the recent review in \citealt{TG09}).

Also note that the linear density field is only accurate on large scales ($\gsim$ 10 Mpc at $z\sim7$).  Thus care should be taken in applying tools which rely on the linear density field (such as the standard excursion set formalism) at smaller scales.  Nevertheless, we include in 21cmFAST a feature to evolve the density field using fully linear evolution, instead of the perturbation approach discussed above.  This allows one to generate extremely large boxes at low resolution.  When using this feature, one should make sure that the chosen cell size is indeed in the linear regime at the redshift of interest. Some results making use of this feature are presented below.

In Fig. \ref{fig:filter_den_pdfs}, we show the PDFs of the density fields smoothed on scale $R_{\rm filter}$, computed from the gas ({\it solid red curves}), DM ({\it dotted green curves}), and 21cmFAST ({\it dashed blue curves}) fields, at $z=$ 20, 15, 10, 7 ({\it top to bottom}).  From the left panel ($R_{\rm filter}=0.5$ Mpc), we see that as structure formation progresses, we tend to increasingly over-predict the abundance of small scale underdensities, and under-predict the abundance of large-scale overdensities.  However, even at $z=7$, our PDFs are accurate at the percent level to over a dex around the mean density.  Understandably, the agreement between the PDFs becomes better with increasing scale (see the right panel corresponding to $R_{\rm filter}=5$ Mpc).
Interestingly, the DM distributions match the gas quite well, although this is somewhat of a coincidence, as we shall see from the power spectra below.

\begin{figure}
\vspace{+0\baselineskip}
\myputfigure{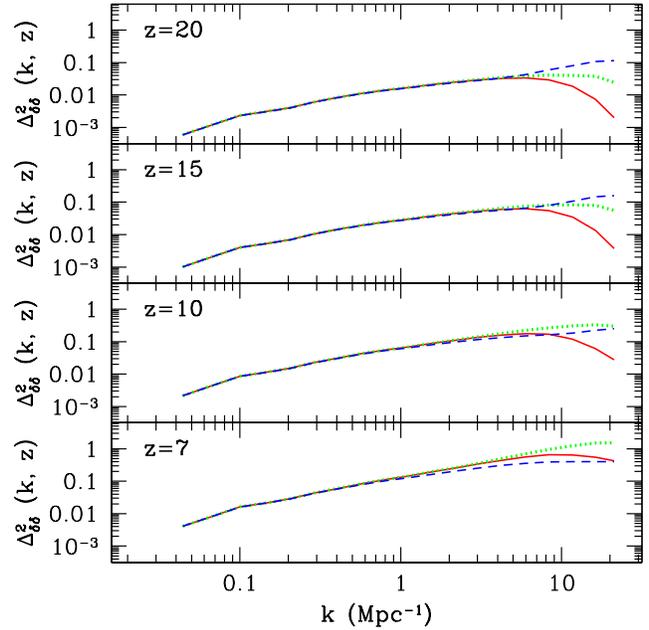}{3.3}{0.5}{.}{0.}
\caption{
Density power spectra, $\denps$, of the gas ({\it solid red curves}), DM ({\it dotted green curves}), and 21cmFAST ({\it dashed blue curves}) fields, at at $z=$ 20, 15, 10, 7 ({\it top to bottom}).
\label{fig:den_ps}}
\vspace{-1\baselineskip}
\end{figure}

In Figure \ref{fig:den_ps}, we present the density power spectra, defined as $\Delta^2_{\rm \delta\delta}(k, z) = k^3/(2\pi^2 V) ~ \langle|\delta({\bf k}, z)|^2\rangle_k$.  The solid red, dotted green, and dashed blue curves correspond to the gas, DM, and 21cmFAST fields, respectively.  On small scales ($k \gsim 5$ Mpc$^{-1}$), the three fields have different power.  The collapse of gas is initially delayed with respect to the pressureless DM, resulting in less small scale power.  The perturbation theory approach of 21cmFAST is closer in spirit to the DM evolution, but does not capture virialized structure.  In fact the close agreement at $z=7$ between the gas and 21cmFAST density power spectra is a coincidence, with the small-scale flattening of the 21cmFAST power attributable to ``shell-crossing'' by the matter particles in the Zel'Dovich approximation.  During reionization, the evolution of the gas is very complicated, since the power spectrum on small scales is sensitive to the thermal history of the reionization model (e.g. \citealt{HG97}).

On large scales ($k\lsim0.5$ Mpc$^{-1}$), all three power spectra agree remarkably at all epochs.  To put this into perspective, neither the MWA nor LOFAR have sufficient signal to noise to detect the 21-cm signal beyond $k\gsim2$ Mpc$^{-1}$ (e.g. \citealt{McQuinn06}).

\subsection{Ionization Field}
\label{sec:ion}

We use a new, refined, semi-numeric algorithm, FFRT, presented in \citet{Zahn10} to generate ionization fields, $\nf({\bf x}, z)$.  The ionization fields have been exhaustively compared against cosmological RT codes in \citet{Zahn10}, yielding good agreement across a broad range of statistical diagnostics on moderate to large scales.  Thus, we will not present further tests here. Instead, we merely outline the procedure, and motivate some aspects with regards to the goals of 21cmFAST: speed and efficiency.

We use the excursion-set approach for identifying HII regions, pioneered by \citet{FZH04}.  The foundation of this approach is to require that the number of ionizing photons inside a region be larger than the number of hydrogen atoms it contains.  Then ionized regions are identified via an excursion-set approach, starting at large scales and progressing to small scales, analogous to the derivation of the Press-Schechter (PS) mass functions \citep{Bond91, LC93}. 

 Specifically, we flag fully ionized cells in our box as those which meet the criteria $\fcoll \geq \zeta^{-1}$, where $\zeta$ is some efficiency parameter and $\fcoll$ is the collapse fraction smoothed on decreasing scales, starting from a maximum $R_{\rm max}$ Mpc and going down to the cell size, $R_{\rm cell}$.  Additionally, we allow for partially-ionized cells by setting the cell's ionized fraction to $\zeta f_{\rm coll}({\bf x}, z, R_{\rm cell})$ at the last filter step for those cells which are not fully ionized\footnote{Our algorithm also can optionally account for Poisson fluctuations in the halo number, when the mean collapse fraction becomes small, $f_{\rm coll}({\bf x}, z, R_{\rm cell}) \times M_{\rm cell} < 50\Mmin$, where $M_{\rm cell}$ is the total mass within the cell and our faintest ionizing sources correspond to a halo mass of $\Mmin$.  This last step is found to be somewhat important when the cell size increases to $\gsim 1$ Mpc (see appendix in \citealt{Zahn10}).  This is left as an option since turning off such stochastic behavior allows the user to better track the deterministic redshift evolution of a single realization.}.
  The ionizing photon horizon, $R_{\rm max}$,  is a free parameter which can be chosen to match the extrapolated ionizing photon mean free path, in the ionized IGM, at $z\sim$ 7--10 (e.g. \citealt{Storrie-Lombardi94, Miralda-Escude03, CFG08}). The photon sinks dominating the mean free path of ionizing photons are likely too small to be resolved in reionization simulations. An effective horizon due to photon sinks can delay the completion of reionization (e.g. \citealt{CHR09, FM09}), and cause a drop in large scale 21-cm power, as we shall see below.

 There are two main differences between FFRT used in 21cmFAST and the previous incarnation of our HII bubble finder used in DexM (MF07): (1) the use of the halo finder to generate ionization fields in DexM (MF07) vs. using the evolved density field and conditional PS to generate ionization fields in 21cmFAST; and (2) the bubble flagging algorithm, which in MF07 is taken to paint the entire spherical region enclosed by the filter as ionized (``flagging-the-entire-sphere''), whereas for 21cmFAST we just flag the central cell as ionized (``flagging-the-central-cell''; for more information, see \citealt{Zahn07, MF07} and the appendix in \citet{Zahn10}. These default settings of 21cmFAST were chosen to maximize speed and dynamic range, while minimizing the memory requirements.  Nevertheless, they are left as user-adjustable options in the codes.

The first difference noted above means that 21cmFAST {\it does not explicitly resolve source halos}.  In MF07, we made use of a semi-numerically generated halo field, which accurately reproduces N-body halo fields down to non-linear scales (MF07; Mesinger et al., in preparation).  As in numerical reionization simulations, these halos were assumed to host ionizing sources.
However, the intermediate step of generating such halo source fields adds additional computation time, and generally requires many GB of RAM for typical cosmological uses.  As for numerical simulations, this memory requirement means that simulation boxes are limited to $\lsim 200$ Mpc if they wish to resolve atomically-cooled halos at $z=$ 7--10, and even smaller sizes if they wish to resolve these at higher redshifts or resolve molecularly-cooled halos. Although DexM's halo finder is much faster than N-body codes, and can generate halo fields at a given redshift in a few hours on a single processor, extending the dynamic range even further without hundreds of GB of RAM would be very useful.  Alternatives to extending the dynamic range have been proposed by \citet{McQuinn07, Santos09}.  These involve stochastically populating cells with halos below the resolution threshold.  Although computationally efficient, it is unclear if these alternatives preserve higher-order statistics of the non-Gaussian $\fcoll$ field, as each cell is treated independently from the others. More fundamentally, the stochasticity involved makes it difficult to deterministically track the redshift evolution of a single realization: halos effectively pop in and out of existence from one redshift output to the next.

Therefore, to increase speed and dynamic range, we use the FFRT algorithm, which uses the conditional PS formalism \citep{LC93, SK99} to generate the collapsed mass (i.e. ionizing source) field.  Since conditional PS operates directly on the density field, without needing to resolve halos, one can have an enormous dynamic range with a relatively small loss in accuracy (compare FFRT and FFRT-S in \citealt{Zahn10}).  Most importantly, when computing $\fcoll$ we use the {\it non-linear} density field, generated according to \S \ref{sec:den}, instead of the standard linear density field.  The resulting ionization fields are a much better match to RT simulations than those generated from the linear density field (\citealt{Zahn10}; foreshadowed also by the ICs panel in Fig. \ref{fig:den_pics} above).  We normalize the resulting collapsed mass field to match the Sheth-Tormen (ST) mean collapse fraction, which in turn matches numerical simulations (see eq. \ref{eq:fcoll} and associated discussion).

The other major difference is that by default, FFRT in 21cmFAST flags just the central filter cell, instead of the entire sphere enclosed by the filter, as in MF07.  The main motivation for this switch is that the former algorithm is ${\cal O}(N)$, while the later is slower: ${\cal O}(N)$ at $\avenf\sim1$ but approaching ${\cal O}(N^2)$ as $\avenf \rightarrow 0$ .  There are some other minor differences between the FFRT and the ionization scheme in MF07, such as the use of a sharp k-space filter instead of a spherical top-hat, but these have a smaller impact on the resulting ionization maps.

\subsection{Peculiar Velocity  Gradient Field}
\label{sec:dvdr}

\begin{figure*}
\vspace{+0\baselineskip}
{
\includegraphics[width=0.45\textwidth]{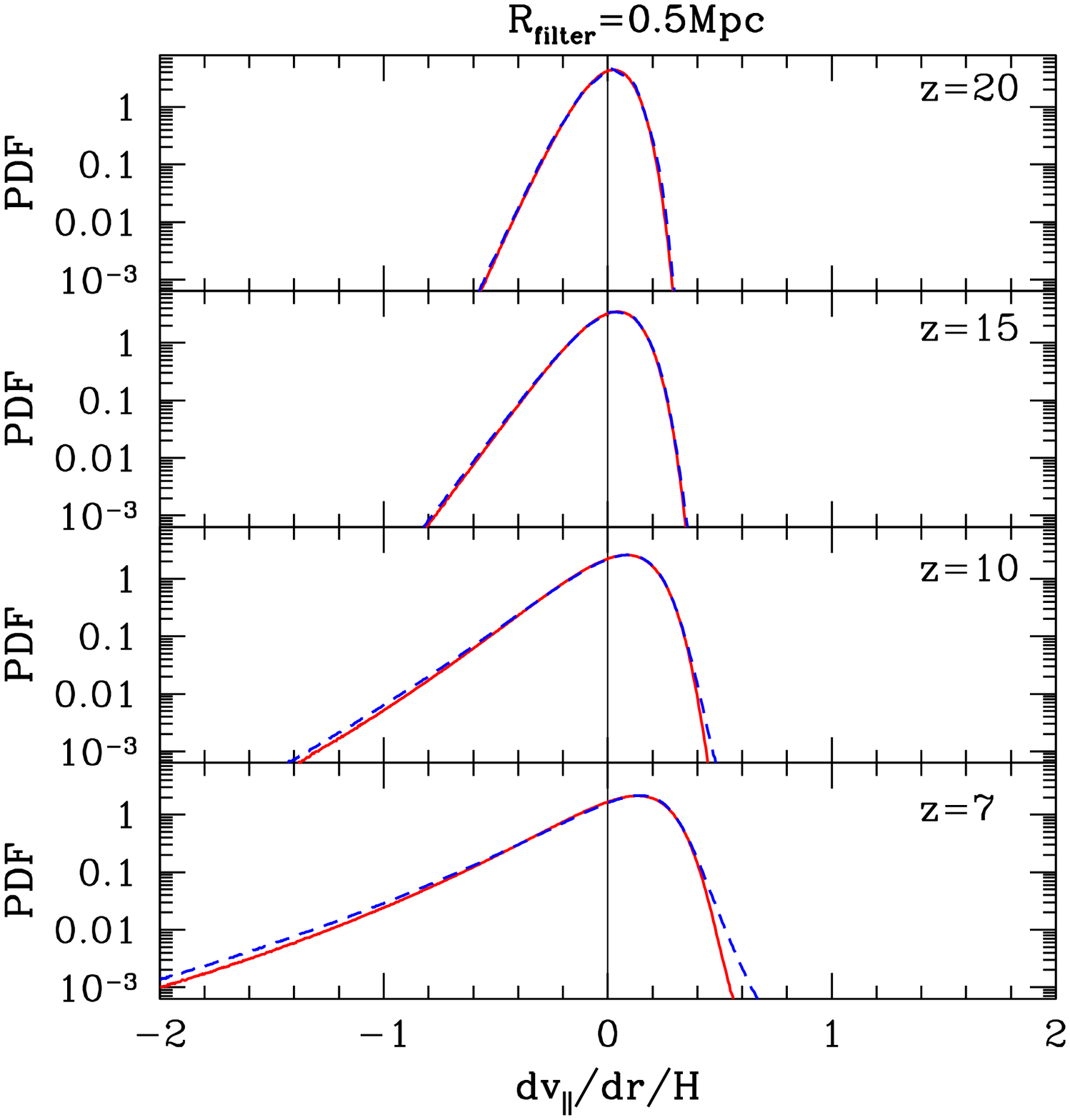}
\includegraphics[width=0.45\textwidth]{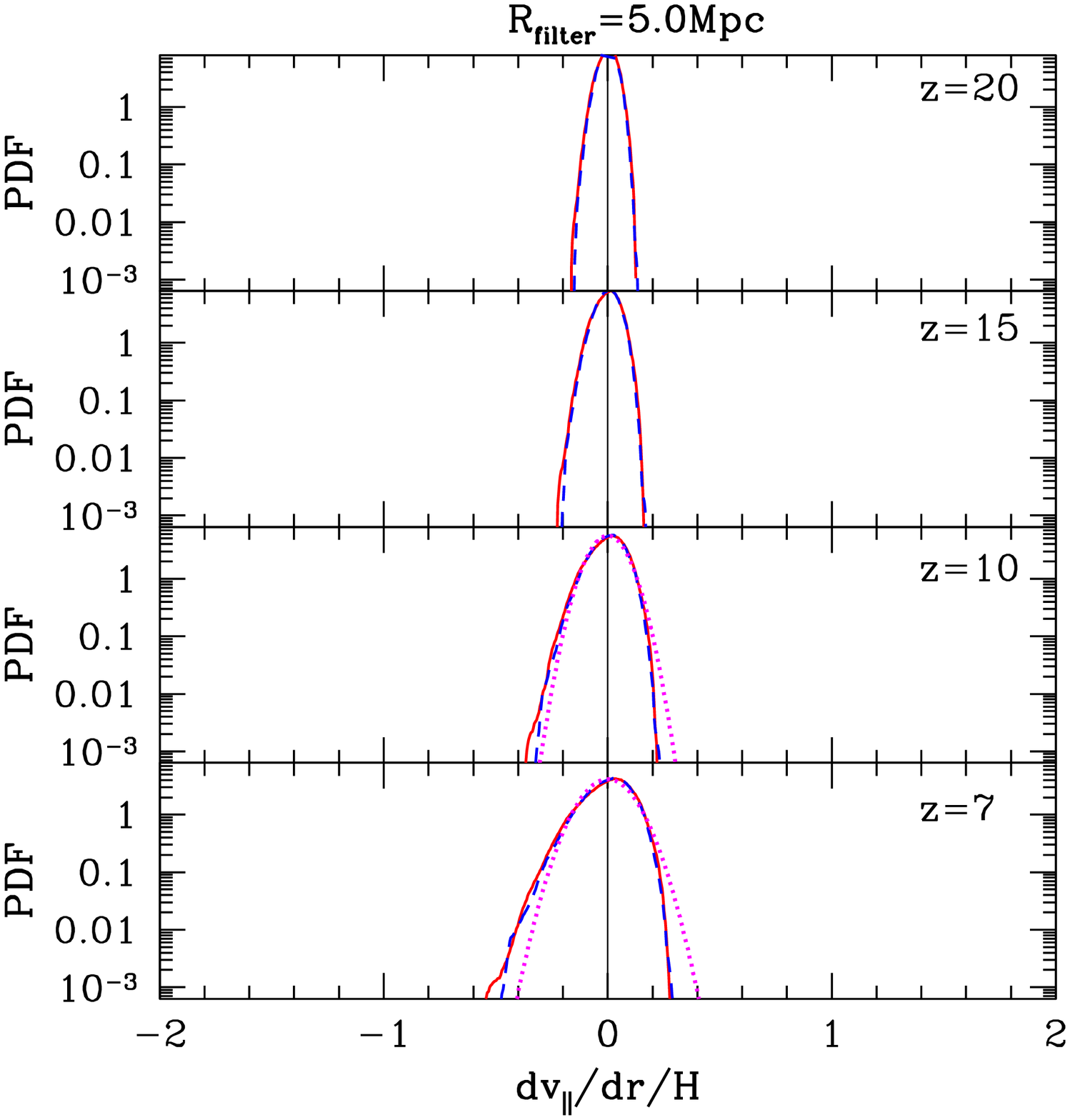}
}
\caption{
PDFs of the comoving LOS derivative of $v_r$ [in units of $H(z)$], smoothed on scale $R_{\rm filter}$ = 0.5 Mpc ({\it left}) and 5.0 Mpc ({\it right}).  Solid red curves are generated from the hydrodynamic simulation, while the dashed blue curves are generated by 21cmFAST.  Redshifts corresponding to $z=$ 20, 15, 10, 7 are shown top to bottom.  All smoothing was performed with a real-space, top-hat filter. The dotted magenta curves were generated on comparable scales by 21cmFAST with different initial conditions; however, they assume linear evolution of the density field, instead of the perturbation theory approach (see \S \ref{sec:den}).
\label{fig:filter_dvdr_pdfs}
}
\vspace{-1\baselineskip}
\end{figure*}

Redshift space distortions, accounted for with the $dv_r/dr$ term in eq. (\ref{eq:delT})\footnote{Note that this expression is exact, as long as the $dv_r/dr$ field is constant over the width of the 21-cm line and $dv_r/dr \ll H(z)$.  Alternately, one can apply redshift space distortions when converting the comoving signal from the simulation box to an observed frequency.  However, for sake of consistency, we perform all of our calculations in comoving space.}, are often ignored when simulating the 21-cm signal.
In the linear regime, redshift space distortions of the 21-cm field are similar to the well-studied Kaiser effect (e.g. \citealt{Kaiser87}), and the power spectrum of fluctuations is enhanced on all scales by a geometric factor of 1.87 \citep{BA04, BL05}.  However, the small scale overdensities, where redshift space distortions are most important, are also the regions whose 21-cm emission is first erased by ``inside-out'' reionization.  Preliminary studies therefore concluded that redshift space distortions would only be noticeable before reionization and in its early stages (\citealt{McQuinn06}, MF07).  As we are interested in accurately simulating the 21-cm signal from all cosmological epochs, including pre-reionization, here we will compare the velocity gradient term from 21cmFAST and hydrodynamic simulations.

Using the Zel'Dovich approximation on our 3D realizations, we can again efficiently move beyond the linear regime into the quasi-linear regime, and take into account correlations in the velocity gradient field.  In this first-order perturbation theory, the velocity field can be written as:

\begin{equation}
\label{eq:velocity}
{\bf v}({\bf k}, z) = \frac{i{\bf k}}{k^2} \dot{D}(z) \delta({\bf k}) ~ ,
\end{equation}

\noindent and so the derivative of the line-of-sight velocity, $v_r$ where ${\bf r}$ for simplicity is oriented along a basis vector, can be written in k-space as:

\begin{align}
\label{eq:dvdr}
\frac{dv_r}{dr}({\bf k}, z) &= i k_r v_r({\bf k}, z)\\
\label{eq:mydvdr}
&\approx - \frac{k_r^2}{k^2} \dot{D}(z) \delta_{\rm nl}({\bf k}) ~ ,
\end{align}

\noindent where differentiation is performed in k-space.  The last approximation is used for 21cmFAST, while the first, exact expression is used for the numerical simulation\footnote{There is an inconsistency in the above equations for 21cmFAST, as eq. (\ref{eq:mydvdr}) is applied to the non-linear density field, whereas eq. (\ref{eq:velocity}) assumes a linear $\delta$.  Nevertheless, as we shall show below, eq. (\ref{eq:mydvdr}) reproduces the non-linear velocity gradient field from the numerical simulations remarkably well.}.

In Fig. \ref{fig:filter_dvdr_pdfs} we show the PDFs of the comoving LOS derivative of $v_r$ [in units of $H(z)$], smoothed on scale $R_{\rm filter}$ = 0.5 Mpc ({\it left}) and 5.0 Mpc ({\it right}).  Solid red curves are generated from the hydrodynamic simulation, while the dashed blue curves are generated by 21cmFAST.  Redshifts corresponding to $z=$ 20, 15, 10, 7 are shown top to bottom. We see that our perturbation theory approach again does remarkably well in reproducing results from the hydrodynamic simulation.  The velocity gradients agree even better than the density fields, since the velocity field is coherent over larger scales. The shape of the distributions are noticeably non-linear on small scales and late times.  The curves resemble PDFs of the sign-flipped non-linear density field, $\delta_{\rm nl}$, which is understandable from eq. (\ref{eq:mydvdr}). The dotted magenta curves in the bottom right panels were generated on comparable scales by 21cmFAST with different initial conditions; however, they assume linear evolution of the density field, instead of the perturbation theory approach.  As expected, linear evolution results in a symmetric Gaussian PDF.

\begin{figure}
\vspace{+0\baselineskip}
\myputfigure{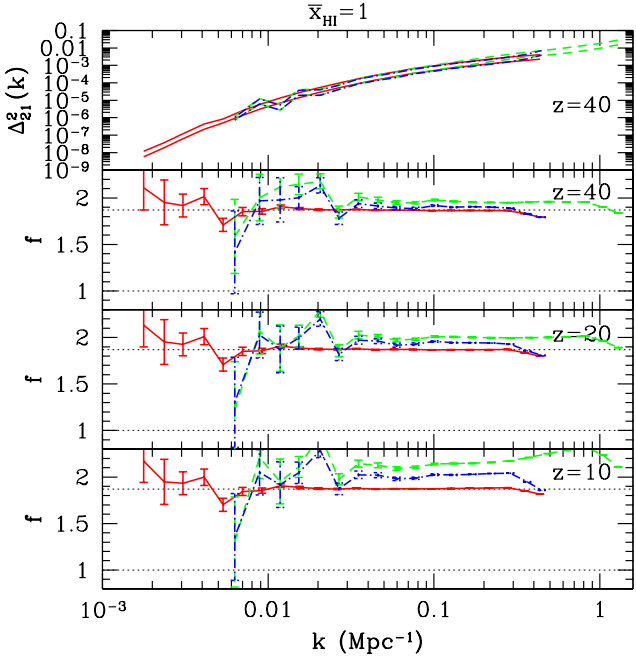}{3.3}{0.5}{.}{0.}
\caption{
{\it Top panel}: Dimensionless power spectra from 21cmFAST, in a fully neutral universe, in the limit of $\Ts \gg \Tcmb$.  The upper set of curves were computed including peculiar velocities, while the lower set were computed not including peculiar velocities.
{\it Bottom three panels}: Ratios of the dimensionless power spectra with peculiar velocities to those not including peculiar velocities. The linear regime geometric enhancement of 1.87 is demarcated by the upper horizontal dotted line.
In all panels, the dot-dashed blue and dashed green curves were generated from the same ICs in a $L=1$ Gpc box, sampled with an initial resolution of $\Delta x$ = 0.56 Mpc, with the evolved density, velocity, and ionization fields generated at lower resolutions of $\Delta x$ = 3.3 Mpc ({\it dashed green curves}), and 10 Mpc ({\it dot-dashed blue curves}).  The solid red curves were generated from a $L=$ 5 Gpc box with a single resolution of $\Delta x$ = 10 Mpc.  However, the density field used for the solid red curves was generated assuming linear evolution, while the others were generated with first-order perturbation theory (see \S \ref{sec:den}).
\label{fig:just_ratios_large_box}}
\vspace{-1\baselineskip}
\end{figure}

\begin{figure}
\vspace{+0\baselineskip}
\myputfigure{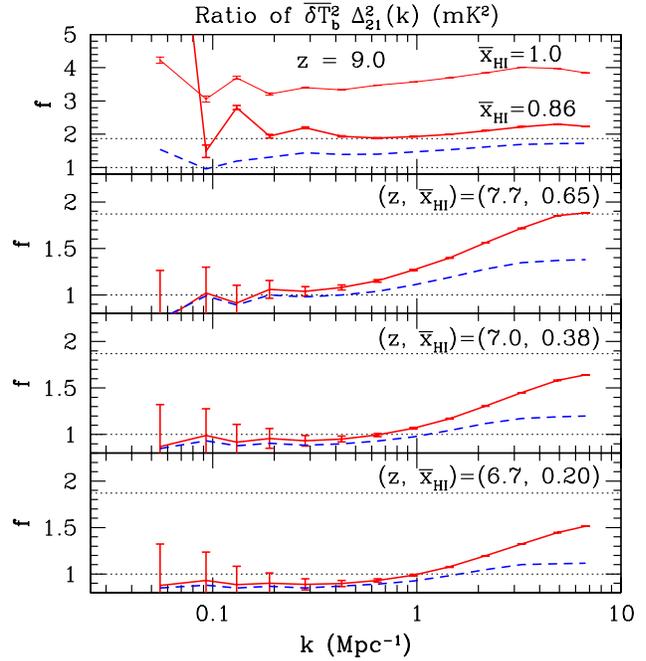}{3.3}{0.5}{.}{0.}
\caption{
Ratios of the {\it dimensional} power spectra, $\bar{\delT}(z)^2 \Delta^2_{21}(k, z)$, computed including peculiar velocities to those not including peculiar velocities.  Panels correspond to $(z, \avenf)$ = (9.00, 0.86), (7.73, 0.65), (7.04, 0.38), and (6.71, 0.20), ({\it top to bottom}).  Solid red curves are generated from the hydrodynamic simulation, while the dashed blue curves are generated by 21cmFAST.  The thin solid red curve in the upper panel corresponds to a fully neutral universe at $z=9$.  The linear regime geometric enhancement of 1.87 is demarcated by the upper horizontal dotted line.
\label{fig:just_ratios}}
\vspace{-1\baselineskip}
\end{figure}

Do we reproduce the geometric, scale-free enhancement of the power spectrum on linear scales?  In the top panel of Fig. \ref{fig:just_ratios_large_box}, we plot {\it dimensionless} 21-cm power spectra, $\Delta^2_{21}(k, z) = k^3/(2\pi^2 V) ~ \langle|\delta_{\rm 21}({\bf k}, z)|^2\rangle_k$ where $\delta_{21}({\bf x}, z) \equiv \delT({\bf x}, z)/ \bar{\delT}(z) - 1$.  The spectra are generated by 21cmFAST in the limit of $\Ts \gg \Tcmb$ and assuming $\avenf=1$.  The solid red curves correspond to a 5 Gpc box with $\Delta x$ = 10 Mpc cells, while the dot-dashed blue and dashed green curves correspond to a 1 Gpc box with different resolutions.  The upper set of curves were computed including peculiar velocities, while the lower set were computed not including peculiar velocities.  The bottom three panels show the ratios of the power spectra that include redshift space distortions to those that do not. 

 Indeed the red curves in Fig. \ref{fig:just_ratios_large_box}, which were evolved linearly, accurately capture the enhancement factor of 1.87, shown with a dotted horizontal line.  The other two curves, which include first order non-linear effects, show an enhancement of power {\it in excess of the purely geometric factor}.  From eq. (\ref{eq:mydvdr}), one sees that a high-value tail in the density distribution resulting from non-linear evolution would drive a corresponding negative tail in the $dv_r/dr$ distributions, which in turn enhances the 21-cm signal through the $(1 / (dv_r/dr/H) + 1)$ term in eq. \ref{eq:delT}. Although the $dv_r/dr$ distributions are zero-mean, the distributions of $1 / (dv_r/dr/H + 1)$ are not.  The bias to higher values is further enhanced when weighted by the local density as in the $\delT$ expression, $\Delta / (dv_r/dr/H + 1)$.  Intuitively, infall in overdense regions causes photons emitted there to travel farther in order to reach a fixed relative redshift; therefore the optical depth and $\delT$ are increased in $\delta > 0$ regions \citep{BL05}.

To further explore this effect and compare our results with simulations, in Fig. \ref{fig:just_ratios}, we plot the ratio of the {\it dimensional} 21-cm power spectra, $\bar{\delT}(z)^2 \Delta^2_{21}(k, z)$, computed including peculiar velocities to those not including peculiar velocities.  The thin 
solid red curve in the upper panel corresponds to a fully neutral universe at $z=9$.  The hydrodynamic simulations go down to much smaller scales than plotted in Fig. \ref{fig:just_ratios_large_box}, which are more non-linear and hence show a larger enhancement of power, though we confirm that most of this is due to the evolution in the mean signal, $\bar{\delT}(z)^2$.

This enhanced 21-cm power from non-linear peculiar velocities obviously merits more investigation beyond the scope of this paper.  Therefore we defer further analysis to future work.  We caution however that it is unclear how well we can estimate this enhancement, due to the misuse of the $1/(dv_r/dr/H+1)$ term in eq. (\ref{eq:delT}).  This expression assumes that $dv_r/dr \ll H(z)$ and diverges at $dv_r/dr=-H(z)$.  To compensate for this behavior, we impose a maximum value of $|dv_r/dr|= 0.5 H(z)$, and confirm that our results are only weakly sensitive to this choice in the $\sim$ $0.1 H(z)$ -- $0.7 H(z)$ range. Similar misuses of the mapping from real space to redshift space have already been noted in the context of galaxy surveys (see \citealt{Scoccimarro04} and references therein).
Therefore, if the user is interested in more accurate predictions of the 21-cm signal as observed with 21-cm interferometers, we recommend to turn off the velocity gradient correction in 21cmFAST, and just do redshift space distortions directly from the velocity field as one of the many necessary transformations from a comoving simulation box to a simulated frequency signal (e.g. \citealt{Harker10}, Matejek et al, in preparation).  The discussion below in the remainder of this section should not be sensitive to the above inconsistency in applying redshift-space corrections.

The remaining curves in Fig. \ref{fig:just_ratios} do not artificially set $\avenf=1$, but use the values of $\avenf$ from the numerical simulations.  Solid red curves are generated from the hydrodynamic simulation, while the dashed blue curves are generated by 21cmFAST.  We confirm the results of MF07: that the enhancement of power due to redshift space distortions vanishes in the early stages of reionization, and subsequently only affects small scales.  Again, this is due to the fact that the densest regions driving most of the enhancement are the first ones to be covered-up by HII regions.  In general, the relative enhancement due to peculiar velocity effects is well-predicted by 21cmFAST on moderate to large scales, $k \lsim 1$ Mpc$^{-1}$, but is under-predicted at small scales and in a very neutral universe.  This is attributable in part to the differences in the ionization field.  Our ionization field algorithm, FFRT, does not properly capture ionization fronts and small-scale HII structure \citep{Zahn10}.  Thus statistics such as this one which are sensitive to small-scales are not accurately reproduced.  21cmFAST's under-prediction of the power spectrum enhancement at small scales is likely also attributable in part to the fact that the Nyquist frequency corresponds to larger scales in the 21cmFAST boxes, since these are directly computed on a $256^3$ grid, whereas the RT simulation is smoothed down from a $512^3$ grid.  This means that our shot noise on small scales in higher than the numerical simulation's, and so fractional enhancements in power should be less (see below).

Finally, we note that the curves in the bottom two panels in Fig. \ref{fig:just_ratios} dip below unity at low $k$.  This means that during the final stages of reionization, peculiar velocity effects actually {\it decrease} power on moderate to large scales.  Although not previously noted in the 21-cm literature, this again can be readily understood:  since reionization is ``inside-out'' on large scales, remaining neutral regions will preferentially be underdensities in the late stages.  Therefore as the average $\delta$ of the remaining neutral regions becomes negative on large scales, $dv_r/dr$ becomes preferentially positive, decreasing power through the $1/(dv_r/dr/H + 1)$ term in eq. \ref{eq:delT}.  We confirm that both the mean signal and the large-scale power show this decrease due to peculiar velocities in the advanced stages of reionization, and it appears in both the simulations and 21cmFAST.

\subsection{Full Post-Heating Comparison of 21-cm Emission}
\label{sec:21cm_cmp}

We now combine the terms from eq. (\ref{eq:delT}) to provide a full comparison of 21cmFAST and numerical simulations, {\it with the assumption of $T_S \gg \Tcmb$}.  For the purposes of these comparisons, the two codes only share the same initial density field; the evolved density, velocity and ionization fields for 21cmFAST are all generated self-consistently as explained above.

\begin{figure*}
{
\includegraphics[width=0.3\textwidth]{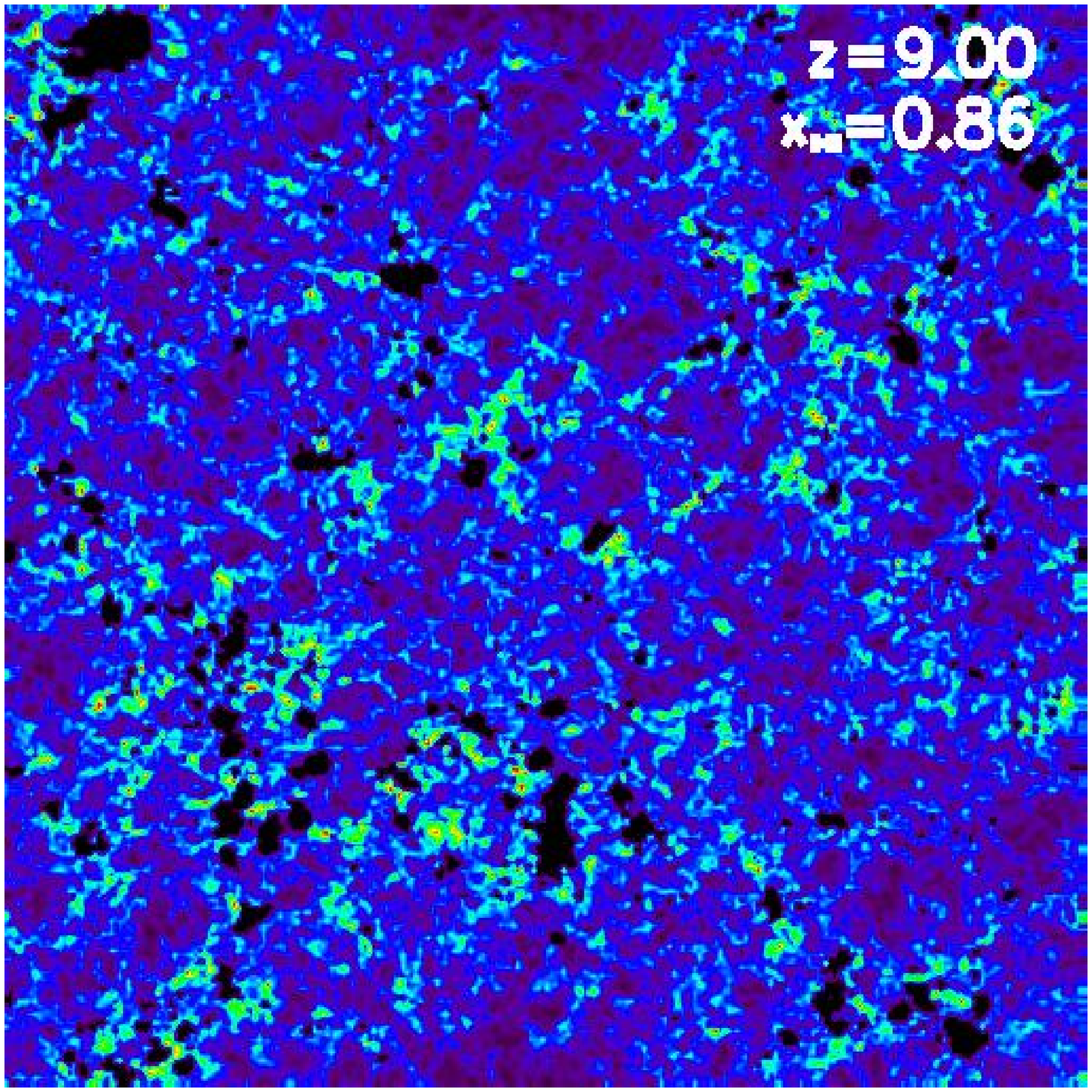}
\includegraphics[width=0.3\textwidth]{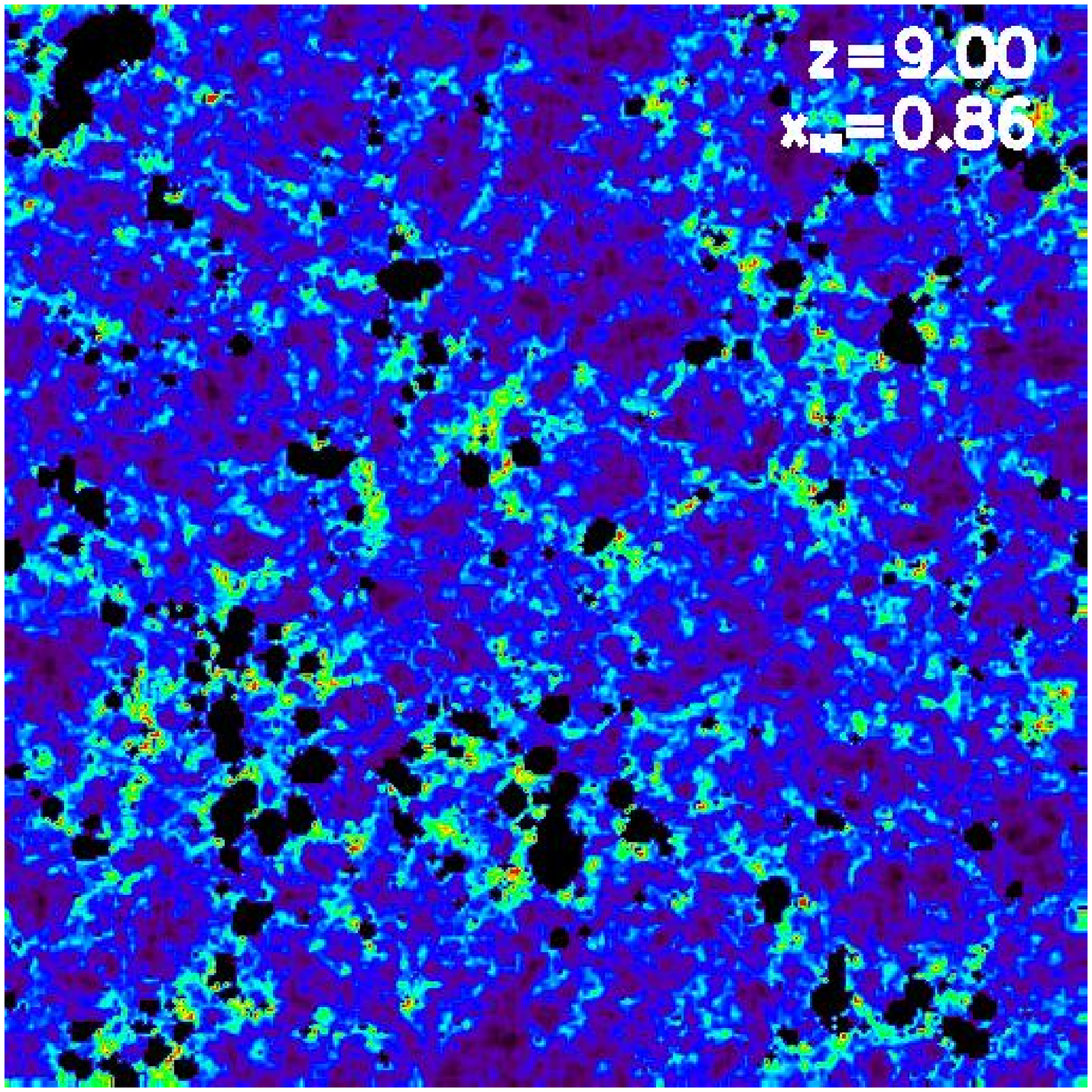}
\includegraphics[width=0.3\textwidth]{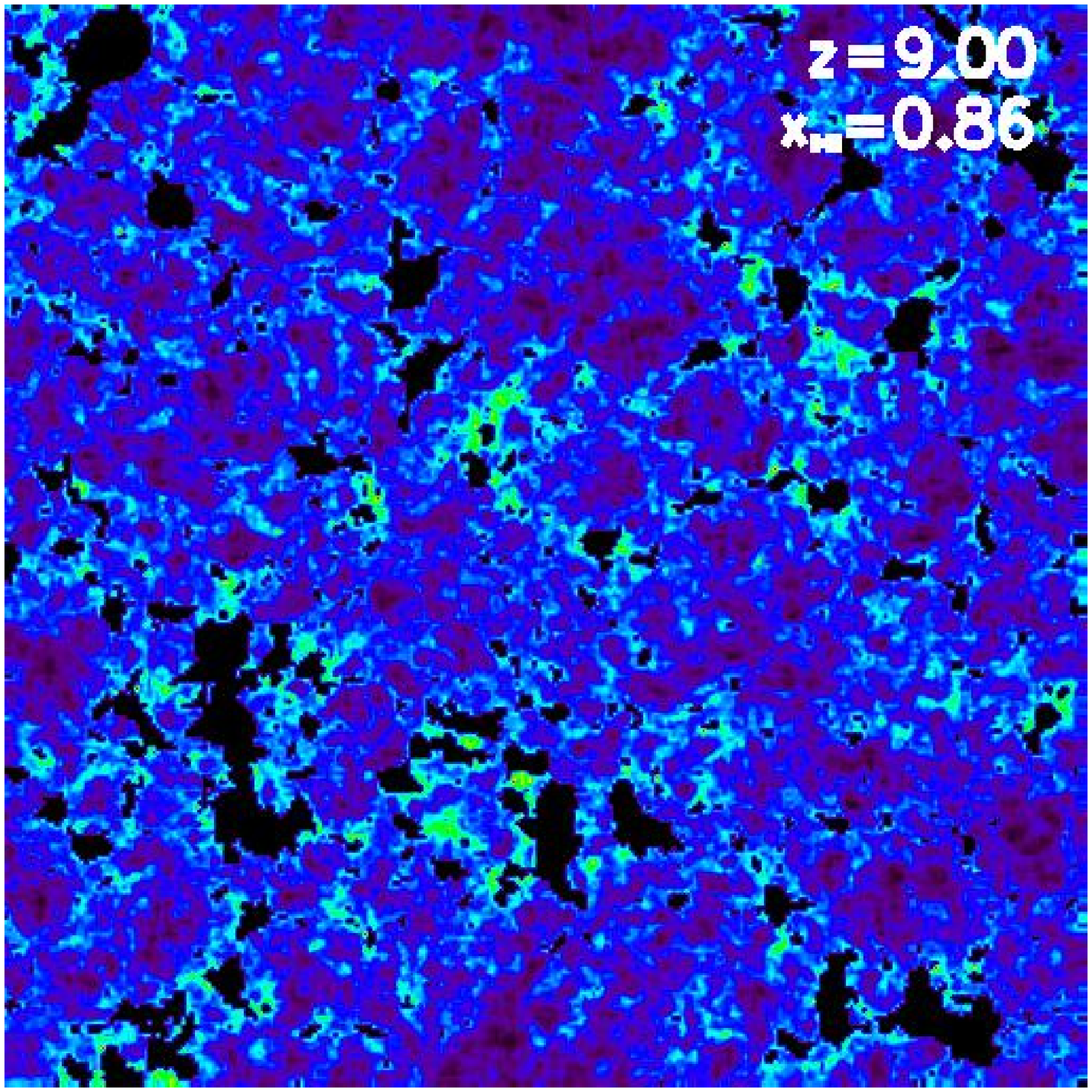}

\includegraphics[width=0.3\textwidth]{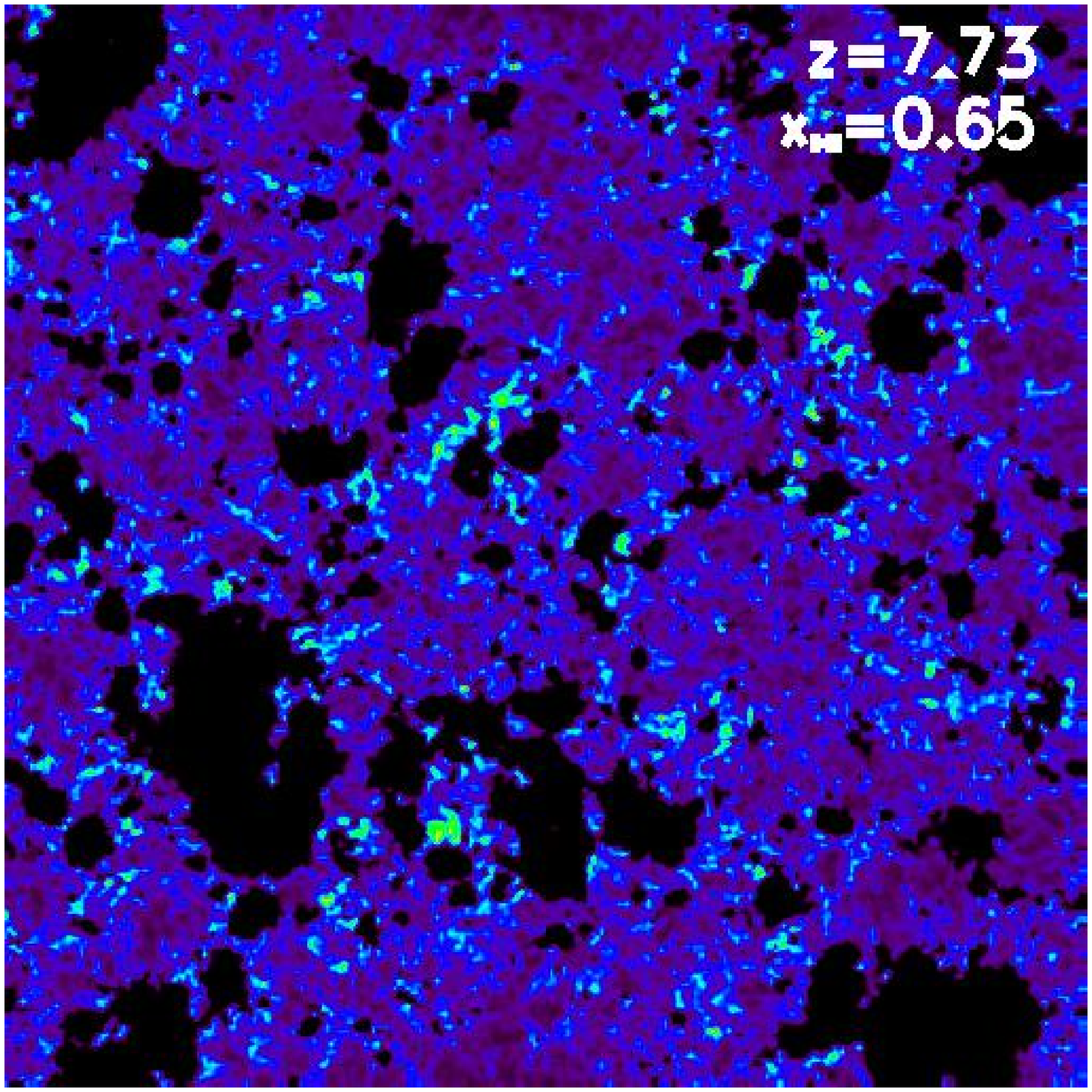}
\includegraphics[width=0.3\textwidth]{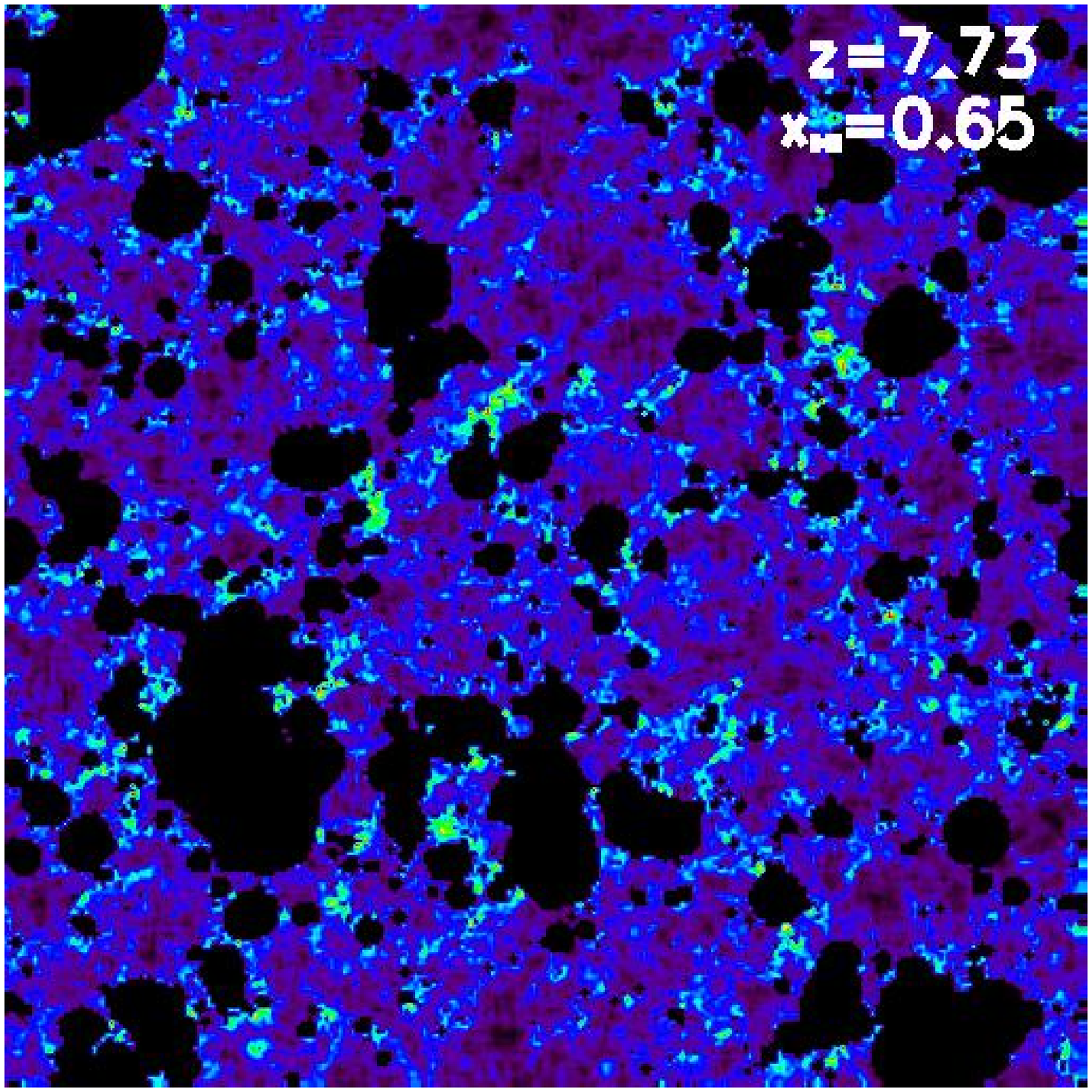}
\includegraphics[width=0.3\textwidth]{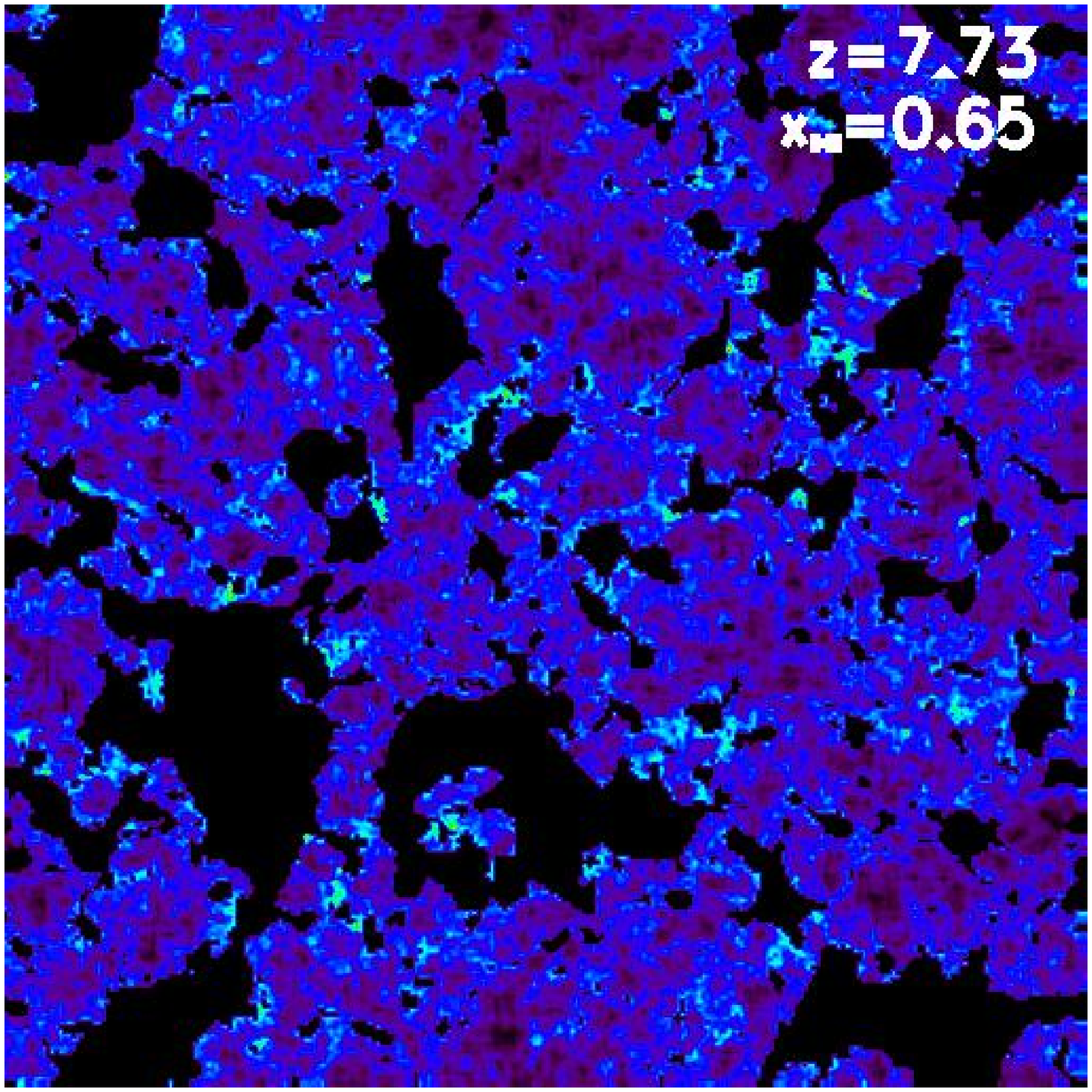}

\includegraphics[width=0.3\textwidth]{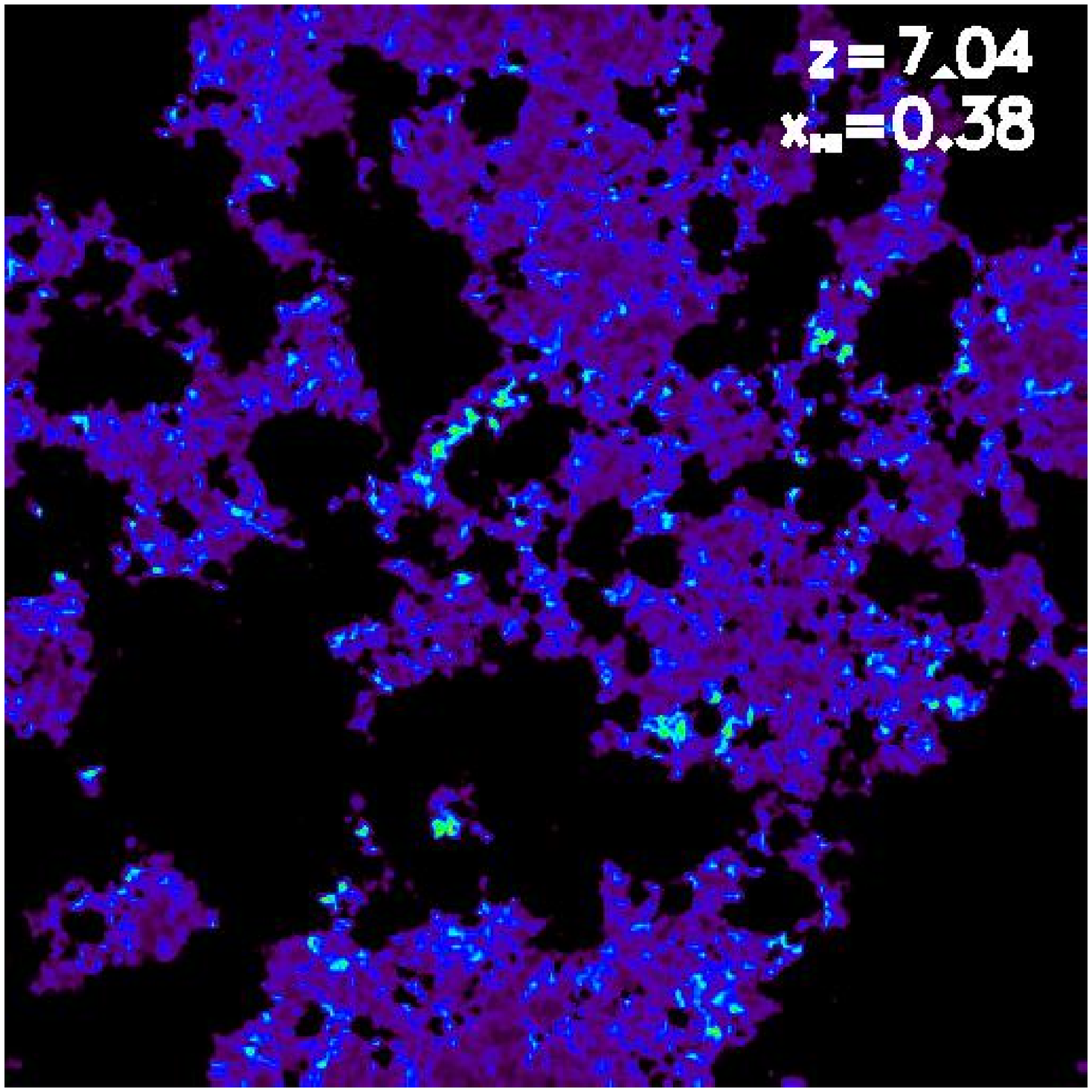}
\includegraphics[width=0.3\textwidth]{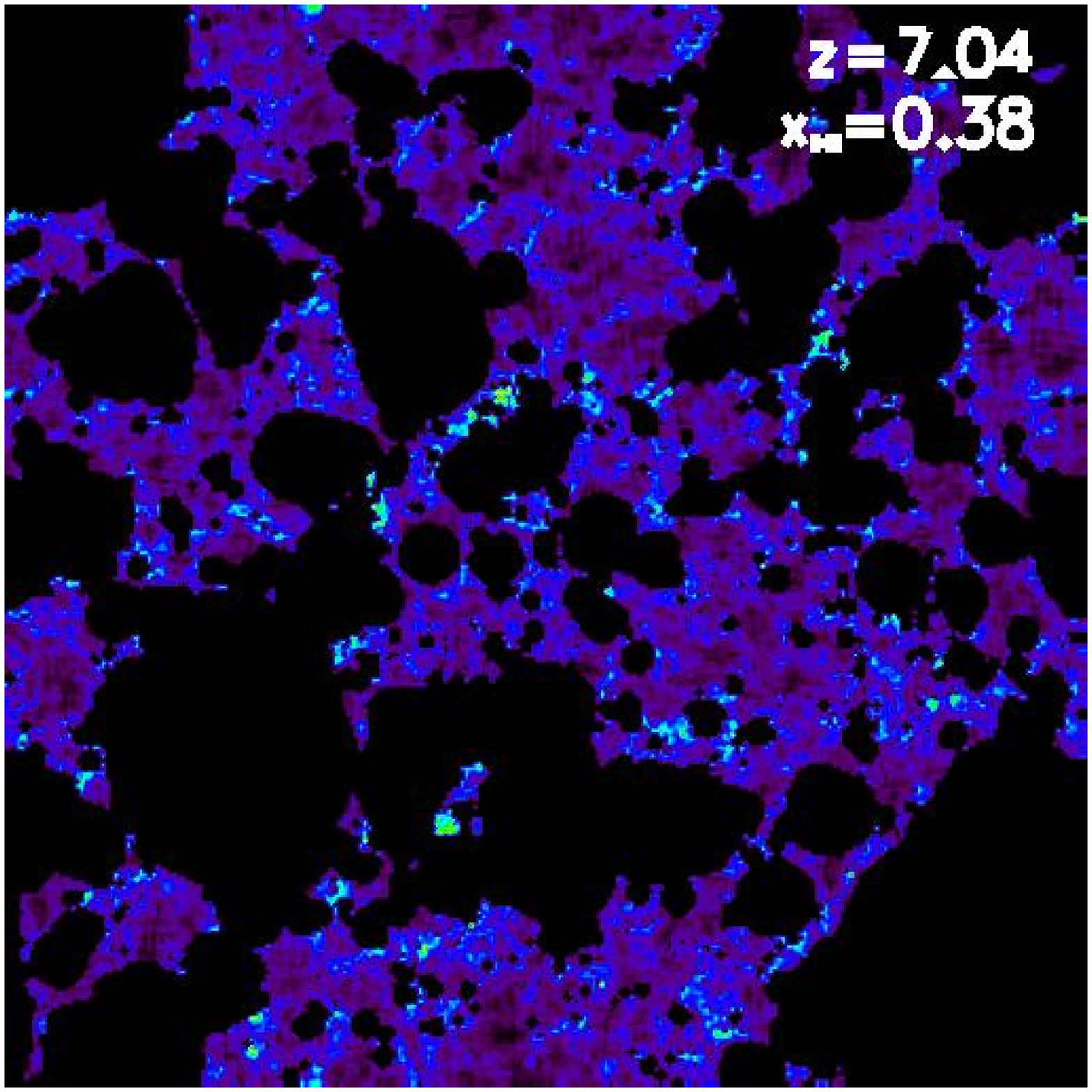}
\includegraphics[width=0.3\textwidth]{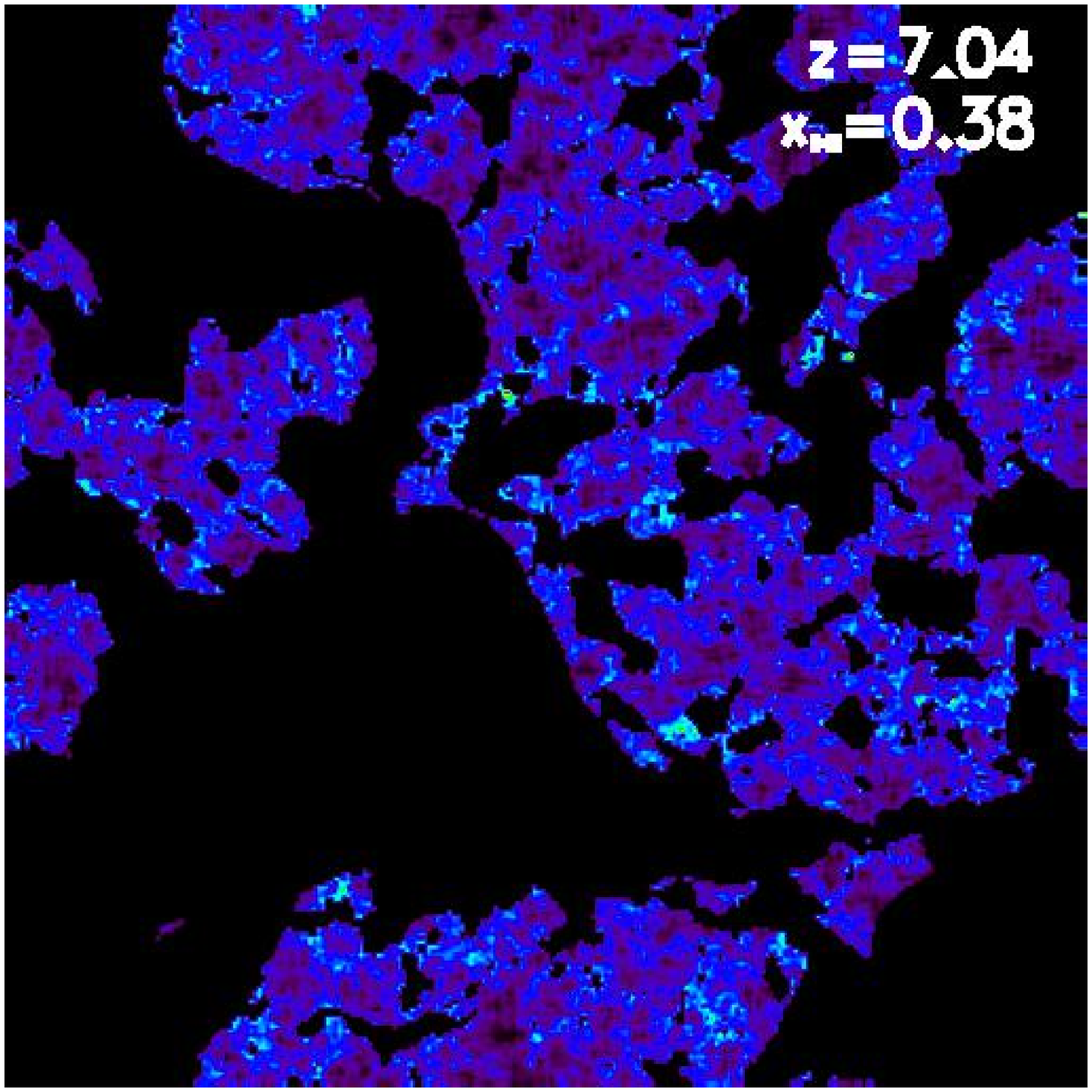}

\includegraphics[width=0.3\textwidth]{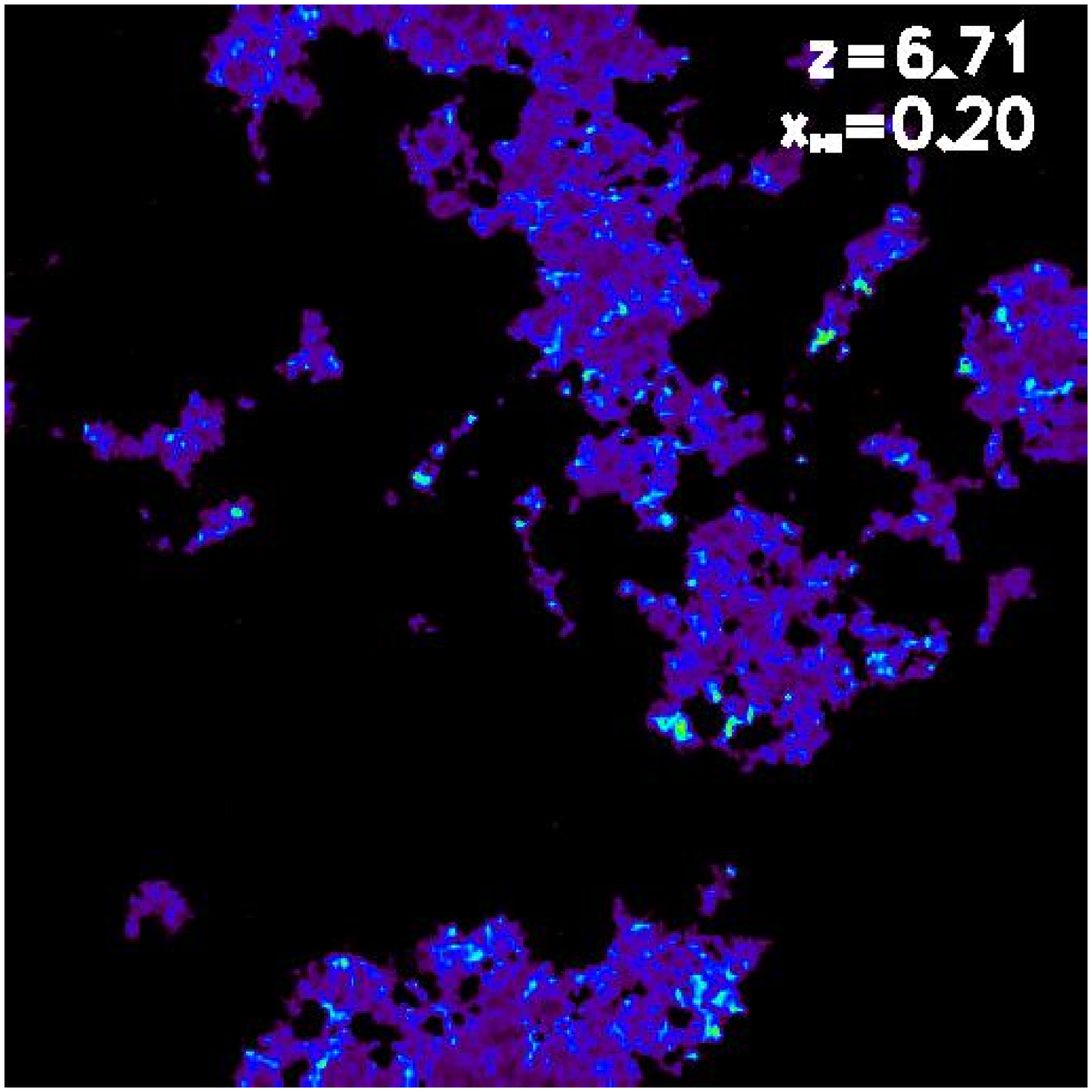}
\includegraphics[width=0.3\textwidth]{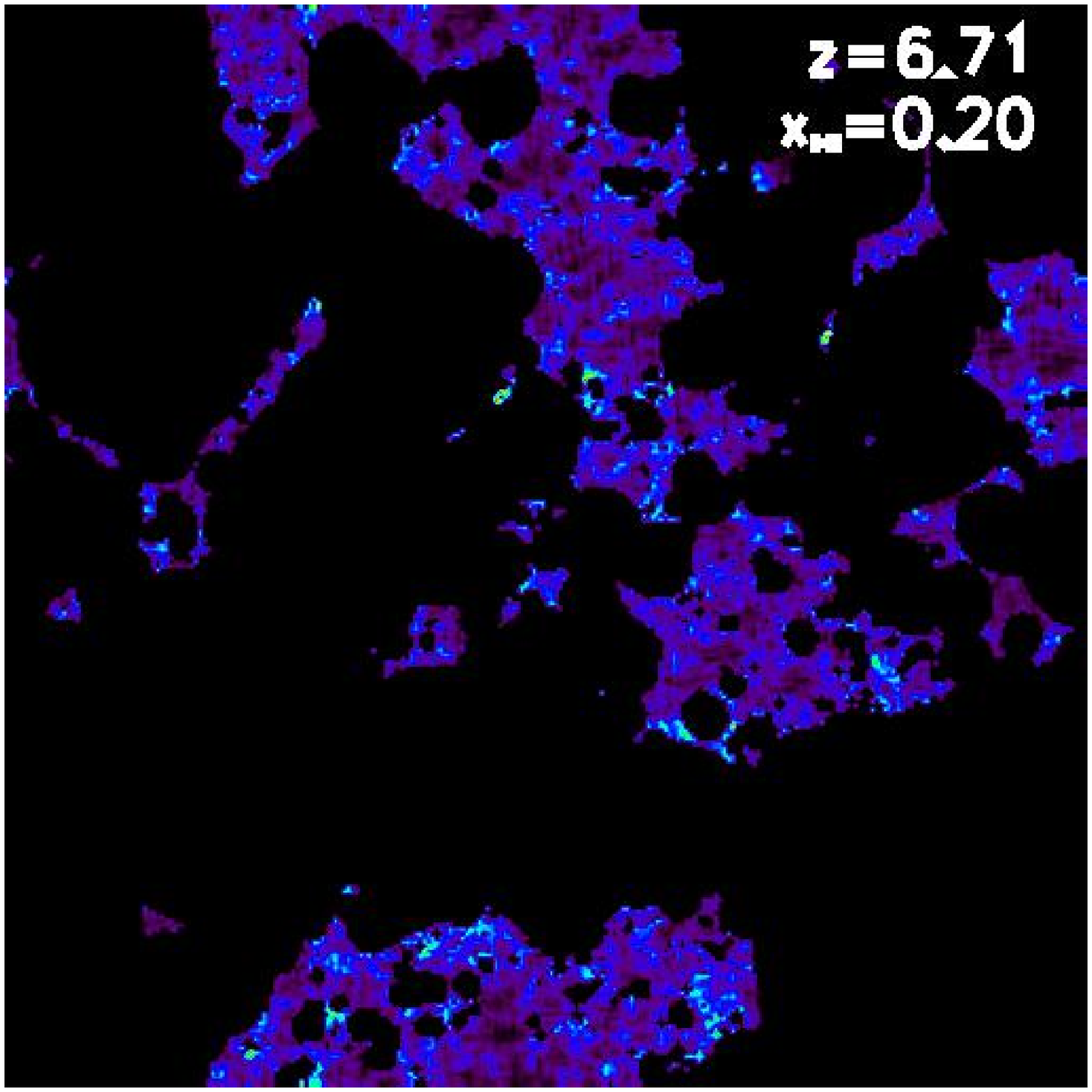}
\includegraphics[width=0.3\textwidth]{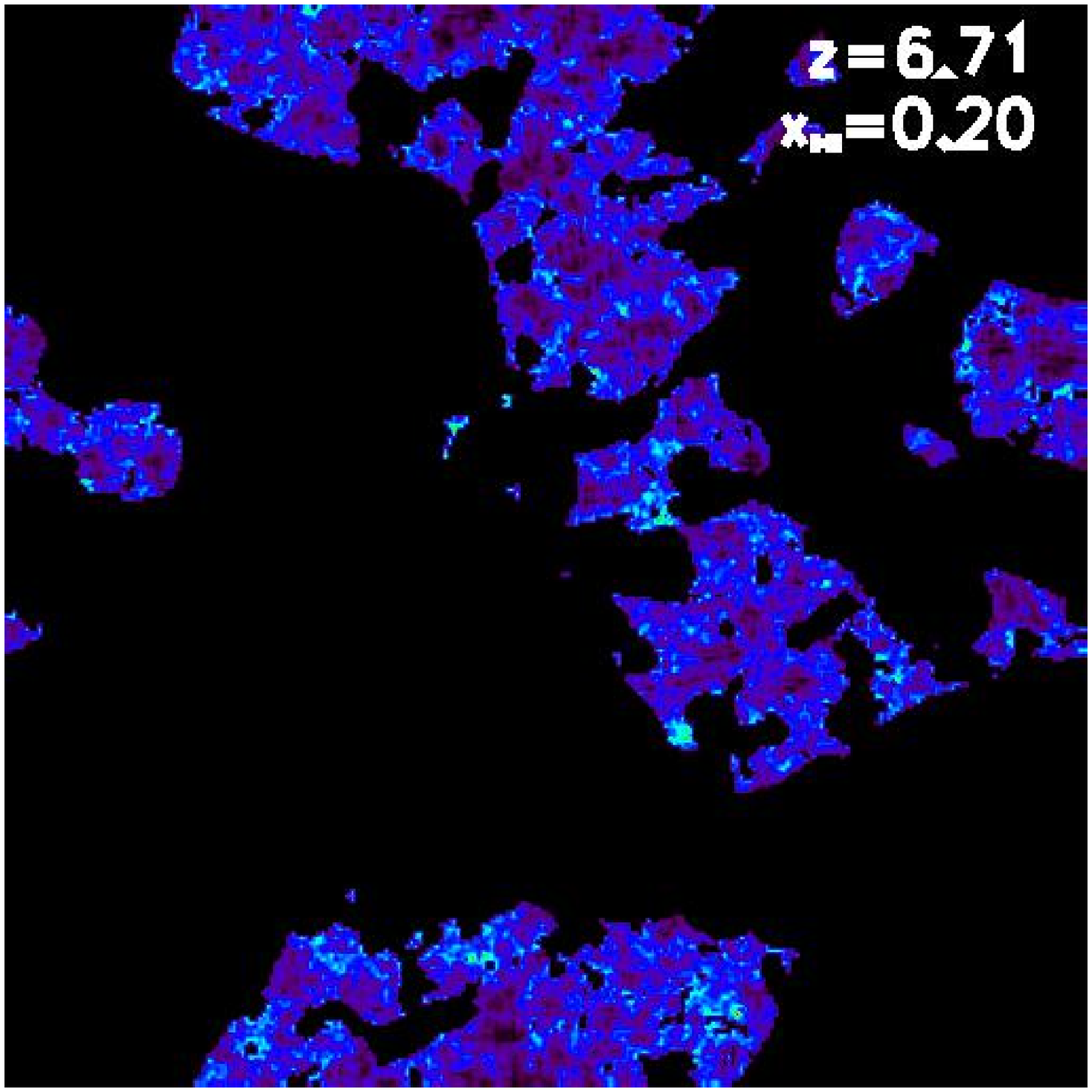}
\vskip0.0pt
}
\includegraphics[width=0.6\textwidth]{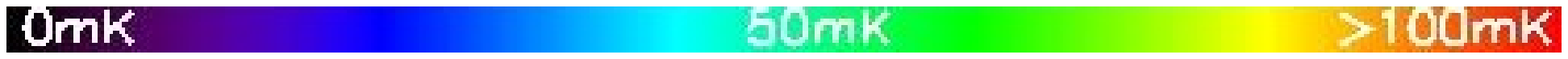}
\caption{
$\delT$ maps. The slices are generated from the hydrodynamic simulation, DexM (MF07), and 21cmFAST, left to right columns.  All slices are 143 Mpc on a side and 0.56 Mpc thick, and correspond to $(z, \avenf)$ = (9.00, 0.86), (7.73, 0.65), (7.04, 0.38), and (6.71, 0.20), top to bottom.
\label{fig:compare_slices}
}
\vspace{-1\baselineskip}
\end{figure*}

\begin{figure*}
\vspace{+0\baselineskip}
{
\includegraphics[width=0.45\textwidth]{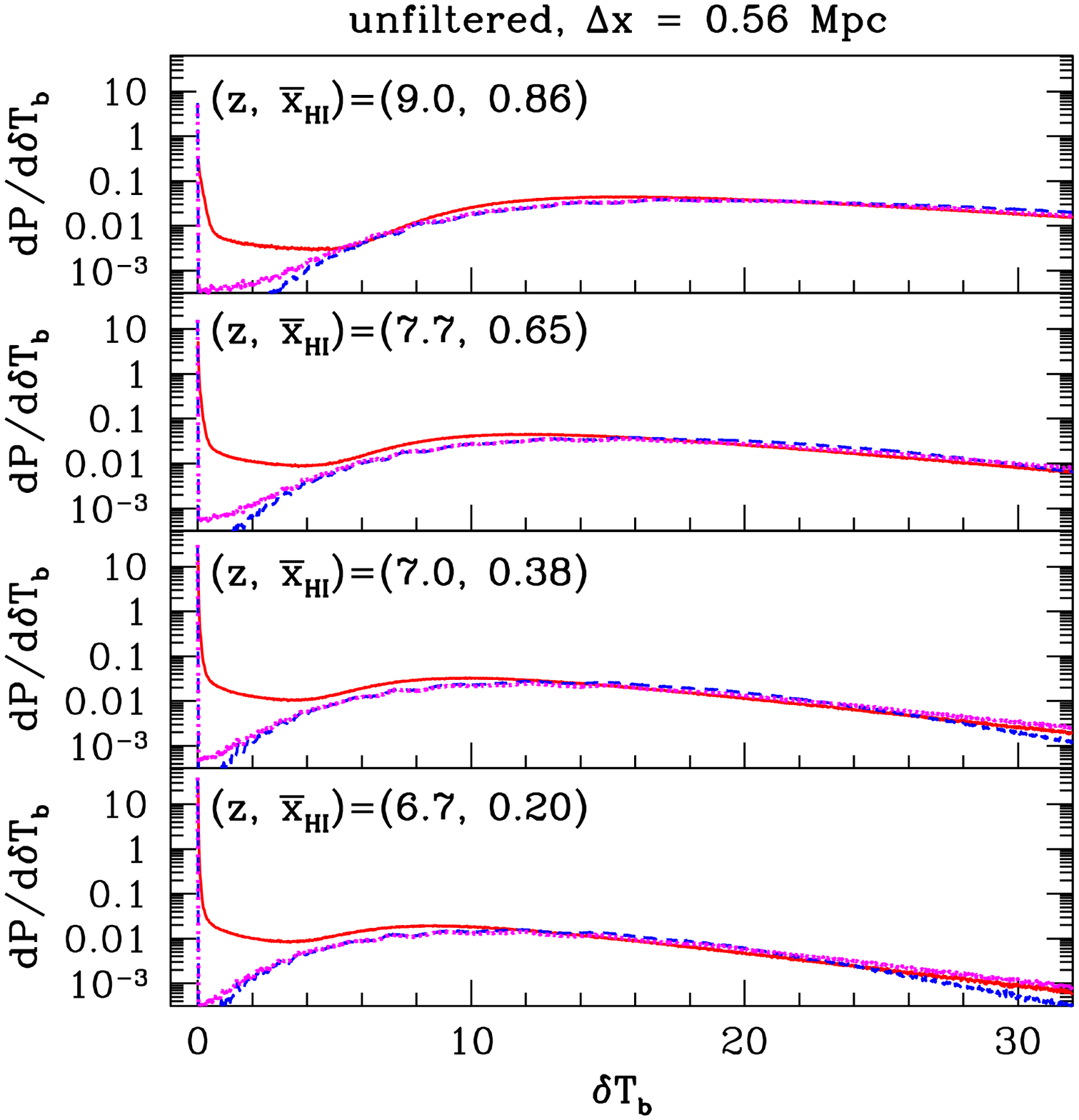}
\includegraphics[width=0.45\textwidth]{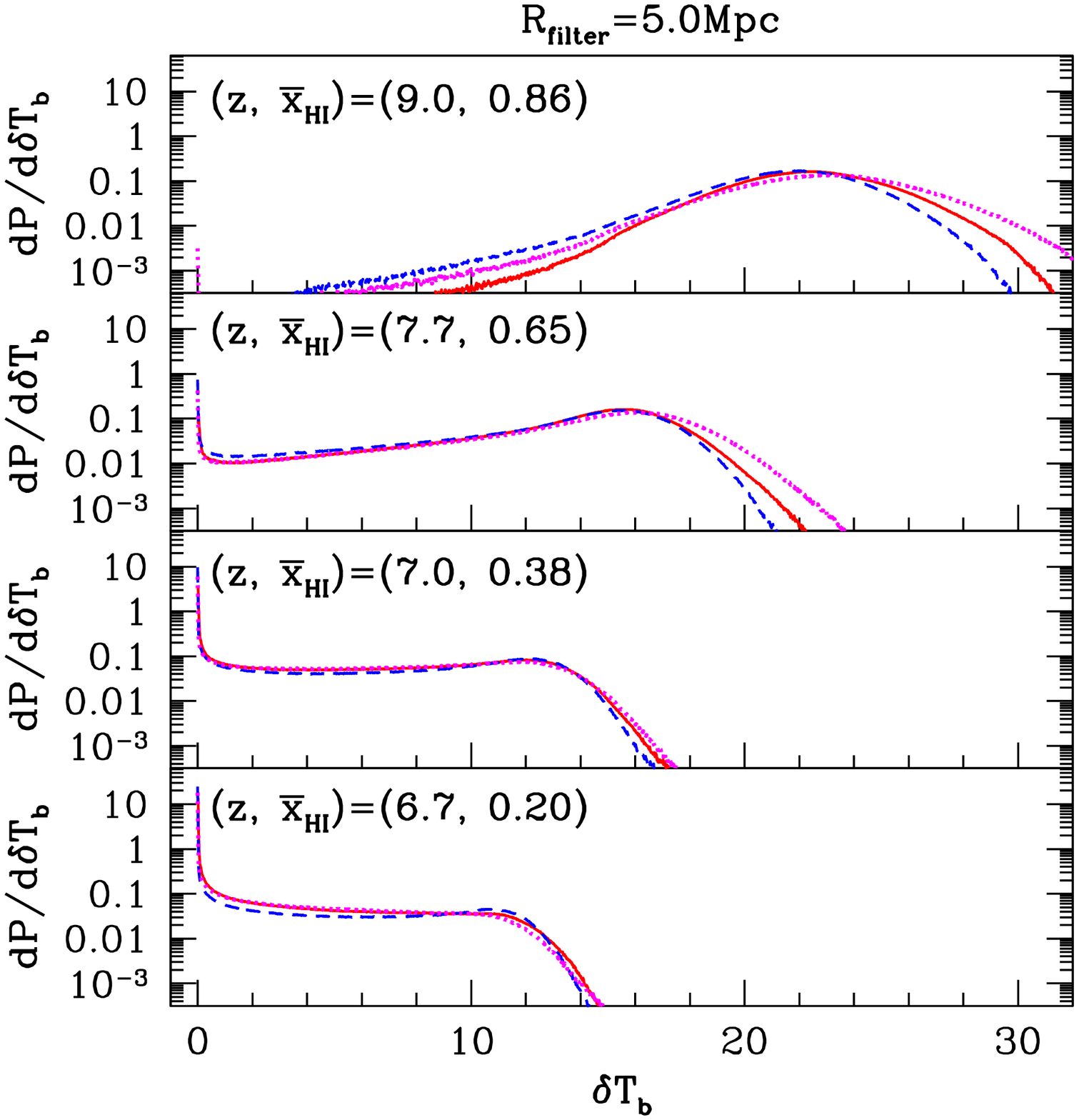}
}
\caption{
PDFs of $\delT$ created using eq. (\ref{eq:delT}) for the hydrodynamic simulation ({\it red solid curves}), 21cmFAST ({\it blue dashed curves}), and MF07 ({\it magenta dotted curves)}.  Panels correspond to  $(z, \avenf)$ = (9.00, 0.86), (7.73, 0.65), (7.04, 0.38), and (6.71, 0.20), top to bottom. The left panel was generated using the unfiltered $\delT$ field with cell length $\Delta x$ = 143/256 Mpc (effectively $R\sim 0.35$ Mpc), while the right panel was generated from the $\delT$ field, filtered on $R_{\rm filter} = 5$ Mpc scales.  All fields invoke the simplifying assumption of $T_S \gg \Tcmb$.
\label{fig:delT_pdfs}
}
\vspace{-1\baselineskip}
\end{figure*}

In Fig. \ref{fig:compare_slices}, we plot slices through the $\delT$ signal, generated from hydrodynamic simulation, the algorithm outlined in MF07\footnote{This algorithm isn't exactly the same as in MF07, since partially ionized cells are still allowed.  However this does not affect our results, since on scales as small as our cell size, the ionization field produced by cosmological RT simulations can be treated as binary, (i.e. either fully ionized or fully neutral; see the Appendix of \citealt{Zahn10}).  The ionization field becomes less binary in the late stages of reionization, or when hard spectra dominate reionization.}, and 21cmFAST, left to right columns.  Rows correspond to $(z, \avenf)$ = (9.00, 0.86), (7.73, 0.65), (7.04, 0.38), and (6.71, 0.20), top to bottom.

As already shown in \S \ref{sec:den}, the density fields are modeled quite accurately with perturbation theory.  One can also see that both semi-numerical schemes reproduce the large-scale HII region morphology (shown in black) of the RT simulations.  Differences emerge at moderate to small scales, with the FFRT ionization algorithm of 21cmFAST generally resulting in HII regions which are too connected.  This difference is mostly attributable to the bubble flagging algorithm; in general the ``flagging-the-entire-sphere'' algorithm of MF07 better reproduces HII morphological structure than the ``flagging-the-central-cell'' algorithm of \citet{Zahn07} (e.g. MF07).

In Fig. \ref{fig:delT_pdfs}, we show the PDFs of $\delT$ for the hydrodynamic simulation ({\it red solid curves}), 21cmFAST ({\it blue dashed curves}), and MF07 ({\it magenta dotted curves)}.  Panels correspond to  $(z, \avenf)$ = (9.00, 0.86), (7.73, 0.65), (7.04, 0.38), and (6.71, 0.20), top to bottom.  The left panel was generated using the unfiltered $\delT$ field with cell length $\Delta x$ = 143/256 Mpc (effectively $R\sim 0.35$ Mpc), while the right panel was generated from the $\delT$ field, filtered on $R_{\rm filter} = 5$ Mpc scales.

From the left panel, we see that we under-predict the number of ``almost'' fully-ionized cells, $\delT \lsim 10$ mK.  This can be traced to our algorithm for determining the partial ionized fraction in the remaining neutral cells.  Our algorithm assumes that cells are partially ionized by sub-grid sources chewing away at their host cell's HI.  Instead, partially ionized cells on these small scales generally correspond to unresolved ionization fronts from non-local sources (see Appendix in \citealt{Zahn10}).  This discrepancy decreases as the cell size increases, since then the fraction of cells which are ionized by sources internal to the cell increases, and the assumption implicit in our FFRT algorithm becomes increasingly accurate.  Aside from this, the distributions in the left panel agree very well.  This should not be surprising, since for comparison the ionization efficiency of the semi-numerical schemes was chosen so that the mean ionized fraction at these epochs agrees with the numerical simulation 
(i.e. the spikes at $\delT=0$ mK match)\footnote{Strictly speaking, the partial ionized cells do impact this calibration, but since most cells are either fully ionized or fully neutral, this is a small effect.}.  The remainder of the signal at $\delT \gsim 10$ mK merely reflects the density distribution of the neutral cells, and we have already demonstrated in \S \ref{sec:den} that our density fields match the hydrodynamic simulation quite well\footnote{The $\delT$ distributions include the additional check of sampling the density field {\it of the remaining neutral cells}, instead of the entire density field studied in \S \ref{sec:den}.  Therefore the close agreement of the $\delT\gsim10$ PDFs contains an additional confirmation of the accuracy of our ionization algorithms.}.

The right panel of Fig. \ref{fig:delT_pdfs} shows the $\delT$ distributions, smoothed on $R_{\rm filter} = 5$ Mpc scales.  As evidenced by the smaller relative spike at $\delT=0$ mK, the ionization fields on these scales are not as binary (i.e. either fully ionized or fully neutral) as those in the left panel.  Thus the PDFs encode more information on the ionization algorithms.  The top panel at $\avenf=0.86$, shows that the predicted distributions of $\delT$ agree well around the mean signal, but the semi-numerical schemes diverge from the RT in the wings.  As mentioned before, the ``flagging-the-entire-sphere'' ionization algorithm from MF07 results in less connected HII regions, and so there are more isolated 5.0 Mpc neutral patches.  This results in an increased number of high-$\delT$ regions. The converse is true for the FFRT ionization scheme which is the fiducial setting of 21cmFAST.  The agreement between the schemes improves as reionization progresses.

\begin{figure}
\vspace{+0\baselineskip}
\myputfigure{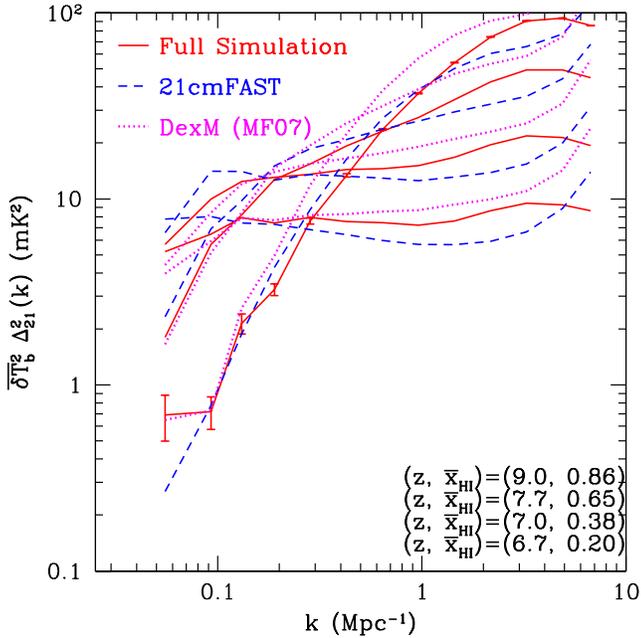}{3.3}{0.5}{.}{0.}
\caption{
Comparison of 21-cm power spectra obtained from the hydrodynamic cosmological simulation ({\it solid red curves}), and the semi-numerical algorithms in MF07 ({\it dotted magenta curves}) and 21cmFAST ({\it dashed blue curves}).  Sets of curves correspond to $(z, \avenf)$ = (9.00, 0.86), (7.73, 0.65), (7.04, 0.38), and (6.71, 0.20), top to bottom at high $k$.
\label{fig:compare_ps}}
\vspace{-1\baselineskip}
\end{figure}

In Fig. \ref{fig:compare_ps}, we compare the power spectra of these $\delT$ boxes.  Again, the hydrodynamic simulation is shown with red solid curves, MF07 is shown with dotted magenta curves, and 21cmFAST is shown with dashed blue curves.

 At all scales, the power spectra agree with each other at the 10s of percent level\footnote{ The semi-numerical simulations show an increase in power approaching the Nyquist frequency, which is most likely just shot noise of our $256^3$ boxes.  The numerical simulations do not show the same upturn, since they were generated on higher resolution grids ($512^3$ for the RT and $768^3$ for the density), and subsequently smoothed down to 256$^3$; numerical simulations generated directly on the same scale $256^3$ grids show similar shot noise upturns in power on these scales (see Fig. 7 in \citealt{Zahn10}).}.
At moderate to large scales, agreement is best, with MF07 performing slightly better than the FFRT algorithm which is default in 21cmFAST \footnote{Note that the FFRT results shown here are not precisely analogous to those in \citet{Zahn10}, since there the evolved density field was taken from an N-body simulation, where in 21cmFAST, we self-consistently generate the density field according to \S \ref{sec:den}.}  On smaller-scales, MF07 predicts too much power, while 21cmFAST under-predicts the power.  It was shown in \citet{Zahn10} that the FFRT ionization algorithm used in 21cmFAST over-predicts the correlation of the ionization and density fields on small scales, due to the fact that it operates directly on the evolved density field.  This strong cross-correlation results in an under-prediction of 21-cm power on these scales.  The converse is true of the MF07 scheme, which although using discrete source halos, paints entire filtered regions as ionized, thus under-predicting the cross-correlation of the ionization and density fields.  The optimal configuration for accurately estimating the 21-cm signal semi-numerically is the FFRT-S scheme discussed in \citet{Zahn10}, set as default in the publicly-available DexM\footnote{http://www.astro.princeton.edu/~mesinger/DexM.html}.

Most importantly, the model uncertainties of the semi-numerical schemes are smaller than the evolution due to reionization over a range $\Delta \avenf \sim 0.2$. Naively therefore, one can predict that the semi-numerical schemes are accurate enough to estimate $\avenf$ from the power spectra to $\pm \lsim 0.1$, or even better if the behavior of the models are understood.  However, there are many astrophysical uncertainties associated with prescriptions for sources and sinks of ionizing photons during the epoch of reionization, and it will likely be these which regulate the achievable constraints on $\avenf$.  {\it Therefore it is imperative for models to be fast and be able to span large regions of parameter space.}  A single 21cmFAST realization of the $\delT$ fields shown in this section (generated from $1536^3$ ICs) takes $\sim$ 30 minutes to compute on a single-processor computer.

\section{The Spin Temperature}
\label{sec:heating}

We now relax the requirement in \S \ref{sec:post_heat}
of $\Ts \gg \Tcmb$, and derive the full 21-cm brightness temperature offset from eq. (\ref{eq:delT}), including the spin temperature field.  As mentioned previously, models predict that the heating epoch concluded well before the bulk of reionization, at $z \gg 10$ \citep{Furlanetto06, CM08, Santos08, Baek09}.  However, the second generation 21-cm interferometers, such as SKA, might be able to peak into this high-redshift regime of the dark ages.  Furthermore, the astrophysical quantities at high-$z$ are uncertain, and we do not really know how robust is the assumption of $\Ts \gg \Tcmb$ even during the early stages of reionization.  Therefore, for many applications, especially parameter studies, it is important to compute the spin temperature field.  Unfortunately, there is currently no numerical simulation which includes the computationally expensive radiative transfer of both X-rays and \lya\ photons from atomically or molecularly cooled sources required to compute $\Ts$ numerically (though see the recent work of \citealt{Baek10}, who perform RT simulations on a small subset of sources, with $M \gsim 10^{10}\Msun$).  Therefore we cannot directly compare our the spin temperature fields to numerical simulations.

 Our derivations in this section are similar to other semi-analytic models \citep{Furlanetto06, PF07, Santos08}.  However, unlike \citet{Santos08} and \citet{Santos09}, we do not explicitly resolve the halo field as an intermediary step.  Instead we operate directly on the evolved density fields, using excursion set formalism to estimate the mean number of sources inside spherical shells corresponding to some higher redshift. As discussed above, bypassing the halo field allows the code to be faster, with modest memory requirements.  Below we go through our formalism in detail.

The spin temperature can be written as (e.g. \citealt{FOB06}):

\begin{equation}
\label{eq:Ts}
\Ts^{-1} = \frac{ \Tcmb^{-1} + x_\alpha T_\alpha^{-1} + x_c \Tk^{-1} }{1 + x_\alpha + x_c}
\end{equation}

\noindent where $\Tk$ is the kinetic temperature of the gas, and $T_\alpha$ is the color temperature, which is closely coupled to the kinetic gas temperature, $T_\alpha \approx \Tk$ \citep{Field59}.  There are two coupling coefficients in the above equation.  The collisional coupling coefficient can be written as:

\begin{equation}
\label{eq:xc}
x_c = \frac{0.0628 ~ \rm K}{A_{10} \Tcmb} \left[ n_{\rm HI} \kappa^{\rm H H}_{1-0}(\Tk) + n_e \kappa^{\rm e H}_{1-0}(\Tk) + n_{\rm p} \kappa^{\rm p H}_{1-0}(\Tk) \right] ~ ,
\end{equation}

\noindent where $A_{10} = 2.85 \times 10^{-15}$ s$^{-1}$ is the spontaneous emission coefficient, $n_{\rm HI}$,  $n_e$, and $n_{\rm p}$ are the number density of neutral hydrogen, free electrons, and protons respectively, and $\kappa^{H H}_{1-0}(\Tk)$,  $\kappa^{e H}_{1-0}(\Tk)$, and $\kappa^{\rm p H}_{1-0}(\Tk)$ are taken from \citet{Zygelman05}, \citet{FF07}, and \citet{FF07}, respectively.  The Wouthuysen-Field (\citealt{Wouthuysen52, Field58}; WF) coupling coefficient can be written as:

\begin{equation}
\label{eq:xa}
x_\alpha = 1.7 \times 10^{11} (1+z)^{-1} S_\alpha J_\alpha ~ ,
\end{equation}

\noindent where $S_\alpha$ is a correction factor of order unity involving detailed atomic physics, and $J_\alpha$ is the Lyman $\alpha$ background flux in units of pcm$^{-2}$ s$^{-1}$ Hz$^{-1}$ sr$^{-1}$.  We compute  $T_\alpha$ and $S_\alpha$ according to \citet{Hirata06}.

According to the above equations, there are two main fields governing the spin temperature: (1) the kinetic temperature of the gas, $\Tk\xz$, and (2) the \lya\ background, $J_\alpha\xz$.  We address these in \S \ref{sec:Tk} and \S \ref{sec:Lya}, respectively.

\subsection{The Kinetic Temperature}
\label{sec:Tk}

\subsubsection{Evolution Equations}
\label{sec:evol}

To calculate the kinetic temperature, one must keep track of the inhomogeneous heating history of the gas.
  We begin by writing down the evolution equation for $\Tk\xz$ and the local ionized fraction in the {\it ``neutral'' (i.e. outside of the ionized regions discussed in \S~\ref{sec:ion})} IGM, $x_e\xz$:

\begin{equation}
\label{eq:ion_rateacc}
\frac{dx_e\xzp}{dz'} = \frac{dt}{dz'} \left[ \Lambda_{\rm ion}
  - \alpha_{\rm A} C x_e^2 n_b f_{\rm H} \right] ~ ,
\end{equation}

\begin{align}
\label{eq:dTkdzacc}
\nonumber \frac{d\Tk\xzp}{dz'} &= \frac{2}{3 k_B (1+x_e)} \frac{dt}{dz'} \sum_p \epsilon_p \\
&+ \frac{2 \Tk}{3 n_b} \frac{dn_b}{dz'}  - \frac{\Tk}{1+x_e} \frac{dx_e}{dz'} ~ ,
\end{align}

\noindent where $n_b=\bar{n}_{b, 0} (1+z')^3 [1+\delNL\xzp]$ is the total (H + He) baryonic number density at $\xzp$, $\epsilon_p\xzp$ is the heating rate {\it per baryon}\footnote{Note that our notation is different than that in \citet{Furlanetto06} and \citet{PF07}, who present heating and ionization rates per proper volume.} for process $p$ in erg s$^{-1}$, $\Lambda_{\rm ion}$ is the ionization rate per baryon, $\alpha_{\rm A}$ is the case-A recombination coefficient, $C\equiv \langle n^2 \rangle / \langle n \rangle^2$ is the clumping factor on the scale of the simulation cell, $k_B$ is the Boltzmann constant, $f_{\rm H}$ is the hydrogen number fraction\footnote{Equation (\ref{eq:ion_rateacc}) ignores Helium recombinations, which is a good approximation given that most He recombining photons will cause ionizations of HI or HeI.}, and we distinguish between the output redshift of interest, $z$, and some arbitrary redshift higher redshift, $z'$.\footnote{For clarity of presentation, we will only explicitly show dependent variables of functions on the left hand side of equations.  Where is is obvious, we also do not explicitly show dependences on $\xzp$.}  We also make the accurate assumption that single ionized helium and hydrogen are ionized to the same degree, $x_e\xzp$, inside the mainly neutral IGM (e.g. \citealt{WL03_preWMAP}).

 In order to speed-up our calculation, we extrapolate the cell's density to higher redshifts assuming linear evolution from $z$ (the desired output redshift at which we compute the non-linear density field with perturbation theory): $\delNL\xzp \approx \delNL\xz D(z')/D(z)$, where $D(z)$ is the linear growth factor.
 In principle, we should follow the non-linear redshift evolution of each cell's density, $\delNL\xzp$. However, linearly extrapolating backwards from $z$ is a good approximation, considering that the majority of cells sized for cosmological simulations are in the linear or quasi-linear regime at very high redshifts where heating is important 
(see, e.g., Fig.~\ref{fig:filter_den_pdfs}) .  Additionally, the evolution of structure, and thus also of heating, is extremely rapid during this epoch, and is dominated by redshifts not much higher than $z$.  Therefore, we can rewrite eq. (\ref{eq:ion_rateacc}) and eq. (\ref{eq:dTkdzacc}) as:

\begin{align}
\label{eq:ion_rate}
&\frac{dx_e\xzp}{dz'} = \frac{dt}{dz'} \Lambda_{\rm ion} \nonumber \\
  &- \frac{dt}{dz'} \alpha_{\rm A} C x_e^2 f_{\rm H} \bar{n}_{b,0} (1+z')^3 [1+\delNL\xz \frac{D(z')}{D(z)}] ~ ,
\end{align}

\begin{align}
\label{eq:dTkdz}
&\frac{d\Tk\xzp}{dz'} = \frac{2}{3 k_B (1+x_e)} \frac{dt}{dz'} \sum_p \epsilon_p \nonumber \\ 
&+ \frac{2 \Tk}{1+z'} + \frac{2 \Tk}{3} \frac{dD(z')/dz'}{D(z)/\delNL\xz + D(z')} - \frac{\Tk}{1+x_e} \frac{dx_e}{dz'} ~ .
\end{align}

\noindent The first term in eq. (\ref{eq:dTkdz}) corresponds to the heat input, which for our purposes is dominated by the heating processes discussed below.  The second term corresponds to the Hubble expansion, the third corresponds to adiabatic heating and cooling from structure formation, and the fourth corresponds to the change in the total number of gas particles due to ionizations.

\subsubsection{Heating and Ionization Rates}
\label{sec:heat}

At very high redshifts, Compton scattering between the CMB photons and the residual free electrons sets $\Tk=\Tcmb$.  After decoupling, the gas temperature evolves through adiabatic cooling,  $\Tk(z') = 2.73 \times (1+z_{\rm dec}) [(1+z')/(1+z_{\rm dec})]^2$, where the decoupling redshift is given by $z_{\rm dec} \approx 137 (\Omb h^2/0.022)^{0.4} - 1$ \citep{Peebles93}.  However, as this is only approximate, we shall use the actual Compton heating rate (e.g. \citealt{SSS00}) in the sum in eq. (\ref{eq:dTkdz}):

\begin{equation}
\label{eq:comp}
\frac{2}{3 k_B(1+x_e)} \epsilon_{\rm comp} = \frac{x_e}{1+f_{\rm He} + x_e} \frac{8 \sigma_{\rm T} u_\gamma}{3 m_e c} (\Tcmb - \Tk) ~ ,
\end{equation}

\noindent where $f_{\rm He}$ is the helium number fraction, $u_\gamma$ is the energy density of the CMB, and $\sigma_{\rm T}$ is the Thomson cross-section. The initial conditions for $x_e$ and $\Tk$ in 21cmFAST can either be provided by the user, or taken from the publicly available code RECFAST\footnote{http://www.astro.ubc.ca/people/scott/recfast.html} \citep{SSS99}.  

We now proceed to outline our procedure for estimating X-ray heating, which is the dominant heating process in this epoch\footnote{There are two other processes that may be important.  The first is shock heating in the IGM:  with such a small temperature, a population of weak shocks could substantially change the thermal history.  This appears to be the case in at least one simulation \citep{Gnedin04}, although other simulations \citep{KMM06} and analytic models \citep{FL04} predict much smaller effects.  The second is any exotic process, such as dark matter decay or annihilation, that produces X-rays or hot electrons \citep{FOP06}.}. {\it We compute the heating rate per particle by summing contributions from sources located in concentric spherical shells around $\xzp$}.  First, we take the standard ansatz in assuming that sources emit photons with a rate proportional to the growth of the total mass fraction inside DM halos, $f_{\rm coll}$.  Note that while this may not be strictly true, we are averaging over large volumes and many sources, and so it is probably a decent assumption in practice.  Thus the total X-ray emission rate (in s$^{-1}$) per redshift interval from luminous sources located between $z''$ and $z''+ dz''$ (where $z'' \geq z'$) can be written as:

\begin{equation}
\label{eq:Ndot}
\frac{d\Ndot}{dz''} = \zeta_X f_\ast \Omega_b \rho_{\rm crit, 0} (1+ \delta^{R''}_{\rm nl})
 \frac{dV}{dz''} 
 \frac{d f_{\rm coll}}{dt} ~ ,
\end{equation}

\noindent where the efficiency, $\zeta_X$, is the number of photons per solar mass in stars and the remaining terms on the RHS correspond to the total star formation rate inside the spherical shell demarcated by $z''$ and $z'' + dz''$. Specifically, $f_\ast$ is the fraction of baryons converted to stars,
 and $dV(z'')$ is the comoving volume element at $z''$ (i.e. volume of the shell). 
  The collapsed fraction is computed according to a hybrid prescription, similar to \citet{BL04, BL08}:

\begin{equation}
\label{eq:fcoll}
f_{\rm coll}({\bf x}, z'', R'', S_{\rm min}) = \frac{\bar{f}_{\rm ST}}{\bar{f}_{\rm PS, nl}} {\rm erfc}\left[ \frac{\delta_c - \delNL^{R''}} {\sqrt{2 [S_{\rm min} - S^{R''}]}} \right] ~ ,
\end{equation}

\noindent where $R''$ is the comoving, null-geodesic distance between $z'$ and $z''$,
$S_{\rm min}$ and $S^{R''}$ are the mass variances on the scales corresponding to the smallest mass sources and $R''$, respectively, $\delNL^{R''}$ is the {\it evolved density}\footnote{Although the collapse criterion for conditional PS was derived using the linear density field, \citet{Zahn10} find that using the evolved density in eq. (\ref{eq:fcoll}) when computing ionization fields yields a better match to radiative transfer simulations.  However, this distinction is less important in the context of computing X-ray flux fields than ionizing flux fields, since the density is more linear on the most pertinent scales $R''$, due to the larger mean free path of X-ray photons.} smoothed on scale $R''$, which we again linearly extrapolate from $z$: $\bar{\delta}^{R''}_{\rm nl}({\bf x}, z'') = \bar{\delta}^{R''}_{\rm nl}({\bf x}, z) D(z'')/D(z)$, $\delta_c$ is the critical linear density corresponding to virialization, and $\bar{f}_{\rm ST}(z'', S_{\rm min})$ is the mean ST collapsed fraction (with the \citealt{Jenkins01} normalization) while $\bar{f}_{\rm PS, nl}(z'', S_{\rm min}, R'')$ is the mean PS collapse fraction, averaged over the evolved density field, $\delNL^{R''}$.
  Therefore the normalization factor, $\bar{f}_{\rm ST} / \bar{f}_{\rm PS, nl}$, assures that the mean collapse fraction matches the ST collapse fraction (in agreement with N-body simulations; e.g. \citealt{Jenkins01, McQuinn07, TC07}), while the fluctuations are sourced by the conditional PS model applied on the evolved density field\footnote{Also note that eq. (\ref{eq:fcoll}) implicitly assumes that the sources are evenly distributed within $R''$, which is the same assumption inherent in the ionization algorithm.  Although the rate of change of the collapse fraction in each shell can be computed from PS, we hesitate to apply this prescription to X-rays, since the ionization algorithm which uses eq. (\ref{eq:fcoll}) has already been rigorously tested and found to agree well with RT simulations of reionization \citep{Zahn10}.}.  There is an implicit assumption in eq. (\ref{eq:Ndot}) that $dD(z'')/dz'' \ll df_{\rm coll}/dz''$, which is quite accurate for all pertinent epochs, as the large-scale density evolves much slower than the exponential growth of the collapsed fraction.

We assume that the X-ray luminosity of sources can be characterized with a power law of the form, $L_e \propto (\nu/\nu_0)^{-\alpha}$, with $\nu_0$ being the lowest X-ray frequency escaping into the IGM.  We can then write the arrival rate [i.e. number of photons s$^{-1}$ Hz$^{-1}$ seen at $\xzp$] of X-ray photons with frequency $\nu$, from sources within $z''$ and $z''+dz''$ as:

\begin{equation}
\label{eq:L_r}
\frac{d\phi_{\rm X}({\bf x}, \nu, z', z'')}{dz''} =  \frac{d\Ndot}{dz''} \alpha \nu_0^{-1} \left(\frac{\nu}{\nu_0} \right)^{-\alpha-1} \left( \frac{1+z''}{1+z'} \right)^{-\alpha-1} {\rm e}^{-\tau_{\rm X}} ~ ,
\end{equation}

\noindent where the last term accounts for IGM attenuation.  For computation efficiency, we compute the IGM X-ray mean free path through the {\it mean}\footnote{Note that our framework for computing the spin temperature starts to break down in the advanced stages of reionization.  When HII regions become sizable, there will be large sightline-to-sightline fluctuations in the X-ray optical depth, which can only be taken into account with approximate radiative transfer.  In our fiducial model (see below), heating has already saturated at $\avenf\gsim0.99$, so this is not a concern.  However, some extreme models might be able to push the heating regime well into the bulk of reionization; we caution the user against over-interpreting the results from 21cmFAST in that regime.} IGM:

\begin{equation}
\label{eq:tau}
\tau_{\rm X}(\nu, z', z'') = \int_{z''}^{z'} d\hat{z} \frac{c dt}{d\hat{z}} \avenf(\hat{z}) \bar{n}(\hat{z}) \tilde{\sigma}(z', \hat{\nu}) ~ ,
\end{equation}

\noindent where the photo-ionization cross-section is weighted over species,$\tilde{\sigma}(z', \hat{\nu}) \equiv f_{\rm H} (1-\bar{x}_e) \sigma_{\rm H} + f_{\rm He} (1-\bar{x}_e) \sigma_{\rm HeI} + f_{\rm He} \bar{x}_e \sigma_{\rm HeII}$ and is evaluated at $\hat{\nu} = \nu (1+\hat{z})/(1+z')$.  In practice, the contribution of HeII can be ignored.  We also remind the reader that $\avenf$ is the volume filling factor of neutral regions during reionization.


Finally, one obtains the X-ray heating rate per baryon, $\epsilon_{\rm X}$ from eq. (\ref{eq:dTkdz}) by integrating over frequency and $z''$:

\begin{align}
\label{eq:eps}
\nonumber \epsilon_{\rm X}\xzp = &\int_{\nu_0}^\infty d\nu 
 \sum_i (h\nu - E^{\rm th}_i)  f_{\rm heat} f_i x_{i}  \sigma_i \\
&\int_{z'}^{\infty} dz'' \frac{d\phi_{\rm X}/dz''}{4 \pi r_p^2} ~ , 
\end{align}

\noindent where $r_p$ is the proper, null-geodesic separation of $z'$ and $z''$, and the frequency integral includes a sum over species, $i=$ HI, HeI, or HeII, in which $f_i$ is the number fraction, $x_i$ is the cell's species' ionization fraction [which for HI and HeI is $(1-x_e)$, and for HeII is $x_e$], $\sigma_i$ the ionization cross-section, and $E^{\rm th}_i$ is the ionization threshold energy of species $i$.
 The factor $f_{\rm heat}[h\nu - E^{\rm th}_i, x_e({\bf x, z'})]$ is defined as the fraction of the {\it electron energy}, $h\nu - E^{\rm th}_i$, deposited as heat.  We use the new results of \citet{FS09}, to compute $f_{\rm heat}$, as well as $f_{\rm ion}$ and $f_{\rm Ly\alpha}$ below.  These fractions take into account the cell's local ionization state, $x_e$, as opposed to the global IGM value, $\bar{x}_e$, used to calculate the optical depth in eq. (\ref{eq:tau}).

Unfortunately, the double integral eq. (\ref{eq:eps}) is slow to evaluate due to the attenuation term, which depends on both redshift and frequency (eq. \ref{eq:tau}).  To speed-up computation, we make the additional approximation that all photons with optical depth, $\tau_{\rm X} \leq 1$ are absorbed, while no photons with optical depth $\tau_{\rm X} > 1$ are absorbed.  Such a step-function attenuation has been shown to yield fairly accurate ionizing photon flux probability distributions (see Fig. 2 in \citealt{MF09}, although comparisons were limited to a single set of parameters).  This approximation allows us to separate the frequency and redshift integrals in eq. (\ref{eq:eps}), removing the exponential attenuation term from $d\phi_{\rm X}/dz''$ and setting the lower bound of the frequency integral to either $\nu_0$ or the frequency corresponding to an optical depth of unity, $\nu_{\tau=1}(\bar{x}_e, z,' z'')$, whichever is larger.  Expanding and grouping the terms, we obtain:

\begin{align}
\label{eq:eps_explicit}
\nonumber \epsilon_{\rm X}\xzp = &\zeta_{\rm X} \alpha c \nu_0^{-1} f_\ast \Omega_b \rho_{\rm crit, 0}(1+z')^{\alpha+1} \\
\nonumber &\int_{\rm Max[\nu_0, \nu_{\tau=1}]}^\infty d\nu \left( \frac{\nu}{\nu_0} \right)^{-\alpha-1}  \sum_i (h\nu - E^{\rm th}_i)  f_{\rm heat} f_i x_{i}  \sigma_i \\
&\int_{z'}^{\infty} dz'' (1+z'')^{-\alpha+2} (1+\bar{\delta}^{R''}_{\rm nl}) \frac{d f_{\rm coll}}{dz''} ~ .
\end{align}

\noindent Now the integrand in both integrals only depends on a single variable, and the entire frequency integral can be treated as a function of $z''$.

Analogously, we can also express the ionization rate per particle in eq. (\ref{eq:ion_rate}) as:

\begin{align}
\label{eq:lam}
\nonumber \Lambda_{\rm ion}\xzp = &\int_{\rm Max[\nu_0, \nu_{\tau=1}]}^\infty d\nu \sum_i f_i x_{i}  \sigma_i F_i \\
&\int_{z'}^{\infty} dz'' \frac{d\phi_{\rm X}/dz''}{4 \pi r_p^2} ~ ,
\end{align}
\begin{align}
\nonumber F_i = \left(h\nu - E^{\rm th}_i \right) \left( 
\frac{f_{\rm ion, HI}}{ E^{\rm th}_{\rm HI}} + 
\frac{f_{\rm ion, HeI}}{ E^{\rm th}_{\rm HeI}} + 
\frac{f_{\rm ion, HeII}}{E^{\rm th}_{\rm HeII}} \right)
 + 1
\end{align}

\noindent where $f_{\rm ion, j}[h\nu - E^{\rm th}_i, x_e({\bf x, z'}), j]$ is now the fraction of the electron's energy going into {\it secondary} ionizations of species $j$, with the unity term inside the sum accounting for the primary ionization of species $i$.

\subsection{The Ly$\alpha$ background}
\label{sec:Lya}

The Lyman $\alpha$ background has two main contributors: X-ray excitation of HI, $J_{\rm \alpha, X}$; and direct stellar emission of photons between \lya\ and the Lyman limit, $J_{\alpha, \ast}$.  The former can easily be related to the X-ray heating rate, assuming that the X-ray energy injection rate is balanced by photons redshifting out of \lya\ resonance \citep{PF07}:

\begin{align}
\label{eq:Jalpha_X}
\nonumber J_{\rm \alpha, X}\xz &= \frac{c n_b}{4 \pi H(z) \nu_\alpha} \int_{z'}^{\infty} dz'' \frac{d\phi_{\rm X}/dz''}{4 \pi r_p^2} \\
&\int_{\rm Max[\nu_0, \nu_{\tau=1}]}^\infty d\nu \sum_i (h\nu - E^{\rm th}_i)  \frac{f_{\rm Ly\alpha}}{h \nu_\alpha} f_i x_{i}  \sigma_i ~ ,
\end{align}

\noindent where  $f_{\rm Ly\alpha}[h\nu - E^{\rm th}_i, x_e({\bf x, z})]$ is the fraction of the electron's energy going into \lya\ photons.

Because of the high resonant optical depth of neutral hydrogen, photons redshifting into any Lyman-$n$ resonance at $\xz$ will be absorbed in the IGM.  They then quickly and locally cascade with a fraction $f_{\rm recycle}(n)$ passing through \lya\ and inducing strong coupling \citep{Hirata06, PF06}.  Therefore, the direct stellar emission component of the \lya\ background (in pcm$^{-2}$ s$^{-1}$ Hz$^{-1}$ sr$^{-1}$) can be estimated with a sum over the Lyman resonance backgrounds (e.g. \citealt{BL05_WF}):

\begin{align}
\label{eq:Jalpha_stars}
J_{\rm \alpha, \ast}\xz &= \sum_{n=2}^{n_{\rm max}} J_\alpha(n, {\bf x}, z) \nonumber \\
&= \sum_{n=2}^{n_{\rm max}} f_{\rm recycle}(n) \int_z^{z_{\rm max}(n)} dz' \frac{1}{4 \pi} \frac{d\phi_\ast^e(\nu'_n, {\bf x}) / dz'}{4 \pi r_p^2}
\end{align}

\noindent , where the emissivity per unit redshift (no. of photons s$^{-1}$ Hz$^{-1}$) is calculated analogously to the X-ray luminosity above:

\begin{equation}
\label{eq:Lya_emiss}
\frac{d\phi_\ast^e(\nu'_n, {\bf x})}{dz'} = \varepsilon(\nu'_n) f_\ast \bar{n}_{\rm b, 0} (1+\bar{\delta}^{R''}_{\rm nl})
\frac{dV}{dz'} \frac{df_{\rm coll}}{dt} ~ .
\end{equation}

\noindent Here $\varepsilon(\nu)$ is the number of photons produced per Hz per stellar baryon, and is evaluated at the emitted (rest frame) frequency:

\begin{equation}
\label{eq:Lya_freq}
\nu'_n = \nu_n \frac{1+z'}{1+z} ~ .
\end{equation}

\noindent The upper limit of the redshift integral in eq. (\ref{eq:Jalpha_stars}) corresponds to the redshift of the next Lyman resonance:

\begin{equation}
\label{eq:Lya_zmax}
1+z_{\rm max}(n) = (1+z) \frac{1-(n+1)^{-2}}{1-n^{-2}} ~ .
\end{equation}

\noindent Following \citet{BL05_WF}, we truncate the sum at $n_{\rm max}=23$, and use their Population II and Population III spectral models for $\varepsilon(\nu)$.  For computational efficiency, one can rearrange the terms in eq. (\ref{eq:Jalpha_stars}), placing the sum over Lyman transitions inside the redshift integral.  Substituting in eq. (\ref{eq:Lya_emiss}) and simplifying, we obtain:

\begin{align}
\label{eq:Jalpha_stars_permute}
J_{\rm \alpha, \ast}\xz = \frac{f_\ast \bar{n}_{\rm b,0} c}{4 \pi} \int_z^{\infty} &dz' (1+z')^3 (1+\bar{\delta}^{R''}_{\rm nl}) \frac{df_{\rm coll}}{dz'} \nonumber \\
&\sum_{n=2}^{n(z')} f_{\rm recycle}(n) \varepsilon(\nu'_n) ~ ,
\end{align}

\noindent where the contribution from the sum over the Lyman transitions is a function of $z'$, and is zero at $z'>z_{\rm max}(n=2)$.

The total Lyman $\alpha$ background is then just the sum of the above components:

\begin{equation}
\label{eq:Jalpha_tot}
J_{\rm \alpha, tot}\xz = J_{\rm \alpha, X}\xz + J_{\rm \alpha, \ast}\xz ~ .
\end{equation}

\noindent In our fiducial model, we do not explicitly take into account other soft-UV sources of \lya\, such as quasars, assuming that these are sub-dominant to the stellar emission.  However, our framework makes it simple to add additional source terms to the integrand of eq. (\ref{eq:Jalpha_stars}), if the user wishes to explore such scenarios (e.g. \citealt{VG09}).

\subsection{Results: complete $\delT$ evolution}
\label{sec:results}

 All of the results in this section are from an $L=$ 1 Gpc simulation, whose ICs are sampled on a 1800$^3$ grid, with the final low-resolution boxes being 300$^3$ (3.33 Mpc cells).  Our fiducial model below assumes $f_\ast=0.1$, $\zeta_X=10^{57} \Msun^{-1}$ ($\sim$ 1 X-ray photon per stellar baryon)\footnote{This number was chosen to match the total X-ray luminosity per unit star formation rate at low redshifts (see \citealt{Furlanetto06} and references therein for details).},  $h\nu_0=200$ eV, $\alpha=1.5$, $T_{\rm vir, min} = 10^4$ K for all sources (X-ray, Lyman $\alpha$ and ionizing), $C=2$, $R_{\rm max}=30$ Mpc, $\zeta_{\rm ion} = 31.5$\footnote{This emissivity was chosen so that the midpoint of reionization is $z\sim10$ and the end is $z\sim7$.} and the stellar emissivity, $\varepsilon$, of Pop II stars from \citet{BL05_WF} normalized to 4400 ionizing photons per stellar baryon.  The free parameters pertaining to the spin temperature evolution were chosen to match those in \citet{Furlanetto06} and \citet{PF07}, to facilitate comparison.  It is trivial to customize the code to add, for example, redshift or halo mass dependences to these free parameters.  The impressive length of the above list of uncertain astrophysical parameters (which itself is only a simplified description of the involved processes) serves well to underscore the need for a fast, portable code, capable of quickly scrolling through parameter space.

We also note that the $\Ts$ calculations outlined in \S \ref{sec:heating} are the slowest part of the 21cmFAST code (as they involve tracking evolution down to the desired redshift), and therefore should only be used in the regime where they are important ($z\gsim17$ in our fiducial model).  For example, generating a $\delT$ box, assuming $\Ts \gg \Tcmb$, on a 300$^3$ grid takes only a few minutes on single processor (depending on the choice of higher resolution for sampling the ICs).  However, including the spin temperature field takes an additional day of computing time.  Nevertheless, once the spin temperature evolution is computed for a given realization at $z$, all of the intermediate outputs at $z' > z$ can be used to compute $\delT$ at those redshifts at no additional computation cost.

\begin{figure}
\vspace{+0\baselineskip}
\myputfigure{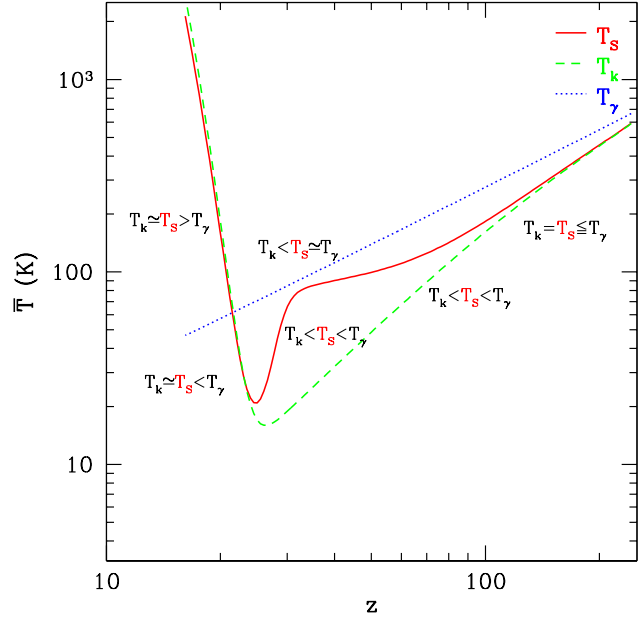}{3.3}{0.5}{.}{0.}
\caption{Evolution of the mean temperatures from 21cmFAST in our fiducial model. Solid, dashed and dotted curves show $\Ts$, $\Tk$ and $\Tcmb$, respectively.
\label{fig:early_global_evolution}}
\vspace{-1\baselineskip}
\end{figure}

Before showing detailed results, it would be useful to summarize the various evolutionary stages (c.f. \S 3.1 in \citealt{PF07}). The reader is encouraged to reference the evolution of the mean temperatures shown in Fig. \ref{fig:early_global_evolution} and/or view the full movie available at http://www.astro.princeton.edu/$\sim$mesinger/Movies/delT.mov, while reading below. 
\begin{packed_enum}

\item {\bf Collisional coupling; $\aveTk = \aveTs \leq \Tcmb$}: At high redshifts, the IGM is dense, so the spin temperature is collisionally coupled to the gas kinetic temperature.  The gas temperature is originally coupled to the CMB, but after decoupling cools adiabatically as $\propto(1+z)^{-2}$, faster than the CMB.  The 21-cm brightness temperature offset from the CMB in this regime starts at zero, when all three temperatures are equal, and then becomes increasingly negative as $\Ts$ and $\Tk$ diverge more and more from $\Tcmb$.  The fluctuations in $\delT$ are driven by the density field, as collisional coupling is efficient everywhere.  In our fiducial model, this epoch corresponds to $100 \lsim z$.

\item {\bf Collisional decoupling; $\aveTk < \aveTs < \Tcmb$}:  The IGM becomes less dense as the Universe expands.  The spin temperature starts to decouple from the kinetic temperature, and begins to approach the CMB temperature again, thus $\delT$ starts rising towards zero.  Decoupling from $\Tk$ occurs as a function of the local gas density, with underdense regions decoupling first.  The power spectrum initially steepens, as small-scale density fluctuations drive the additional fluctuations of the collisional coupling coefficient.  As the spin temperature in even the overdense regions finally decouples from the kinetic temperature, the power spectrum flattens again, and the mean signal drops as 
 $\aveTs \rightarrow 0$.  In our fiducial model, this epoch corresponds to $35 \lsim z \lsim 100$.

\item {\bf Collisional decoupling $\rightarrow$ WF coupling transition; $\aveTk < \aveTs \approx \Tcmb$}: As the spin temperature throughout the IGM decouples from the kinetic temperature, the mean signal is faint and might disappear, if the first sources wait long enough to ignite.  In our fiducial model, this transition regime doesn't really exist.  In fact our first sources turn on before the spin temperature fully decouples from the kinetic temperature.

\item {\bf WF coupling; $\aveTk < \aveTs < \Tcmb$}:  The first astrophysical sources turn on, and begin coupling the spin temperature of the nearby IGM to the kinetic temperature through the WF effect (\lya\ coupling).  As the requirements for \lya\ coupling are more modest than those to heat the gas through X-ray heating, the kinetic temperature keeps decreasing in this epoch.  The mean brightness temperature offset from the CMB starts becoming more negative\footnote{Note that we discuss global trends here.  Locally around each \lya\ source, there are partially ionized regions hosting hotter gas (e.g. \citealt{Cen06}).}
 again and can even reach values of $\delT < -100$ mK. In our fiducial model, this epoch corresponds to $25 \lsim z \lsim 35$.

\item {\bf WF coupling $\rightarrow$ X-ray heating transition; $\aveTk \sim \aveTs < \Tcmb$}: \lya\ coupling begins to saturate as most of the IGM has a spin temperature which is strongly coupled to the kinetic temperature.  The mean spin temperature reaches a minimum value, and then begins increasing. A few underdense voids are left only weakly coupled as X-rays from the first sources begin heating the surrounding gas in earnest, raising its kinetic temperature.  The 21-cm power spectrum steepens dramatically as small-scale overdensities now host hot gas, while on large scales the gas is uniformly cold as \lya\ coupling saturates.  As inhomogeneous X-ray heating continues, the large-scale power comes back up.  In our fiducial model, this transition occurs around $z\sim25$.

\item{\bf X-ray heating; $\aveTk = \aveTs < \Tcmb$}: X-rays start permeating the IGM.  The fluctuations in $\delT$ are now at their maximum, as regions close to X-ray sources are heated above the CMB temperature, $\delT > 0$, while regions far away from sources are still very cold, $\delT < 0$. A ``shoulder'' in the power spectrum, similar to that seen in the epoch of reionization (e.g. \citealt{McQuinn07}), moves from small scales to large scales. X-rays eventually heat the entire IGM, and 21-cm can only be seen in emission.  The power spectrum falls as this process nears completion.  In our fiducial mode, this epoch corresponds to  $18 \lsim z \lsim 25$.

\item {\bf X-ray heating $\rightarrow$ reionization transition; $\aveTk = \aveTs > \Tcmb$}: X-rays have heated all of the IGM to temperatures above the CMB. The 21-cm signal becomes insensitive to the spin temperature.
  Emission in 21-cm is now at its strongest before reionization begins in earnest.  The 21-cm power spectrum is driven by the fluctuations in the density field. In our fiducial model, this epoch corresponds to $16 \lsim z \lsim 18$.

\item {\bf Reionization}:  Ionizing photons from early generations of sources begin permeating the Universe, wiping-out the 21-cm signal inside ionized regions.  The power spectrum initially drops on large scales at $\avenf\gsim0.9$ as the first regions to be ionized are the small-scale overdensities \citep{McQuinn07}.  The mean signal decreases as HII regions grow, and the power spectrum is governed by HII morphology.  This epoch can have other interesting features depending on the detailed evolution of the sources and sinks of ionizing photons, as well as feedback processes, but as the focus of this section is the pre-reionization regime, we shall be brief in this point.  In our fiducial model, this epoch corresponds to $7 \lsim z \lsim 16$.
\end{packed_enum}

These milestones are fairly general, and should appear in most regions of astrophysical parameter space.  However, the details of the signal, as well as the precise timing and duration of the above epochs depends sensitively on uncertain astrophysical parameters.  For example, note that the above epochs in our fiducial model are shifted to higher redshifts than the analogous ones in \citet{Furlanetto06} and \citet{PF07}.  This is because the source abundances in those works were computed with PS, which under-predicts the abundances of $\Tvir > 10^4$ K halos by over an order of magnitude at high redshifts (e.g. \citealt{TC07}).  A similar effect is seen when compared to the recent numerical simulations of \citet{Baek10}, who were only able to resolve halos with mass $>10^{10} \Msun$, which is 2--3 orders of magnitude away from the atomic cooling threshold at these redshifts.  With such rare sources, they heat the gas in their 100$h^{-1}$ Mpc boxes to above the CMB much later, at $z<9$.

\begin{figure}
\vspace{+0\baselineskip}
\myputfigure{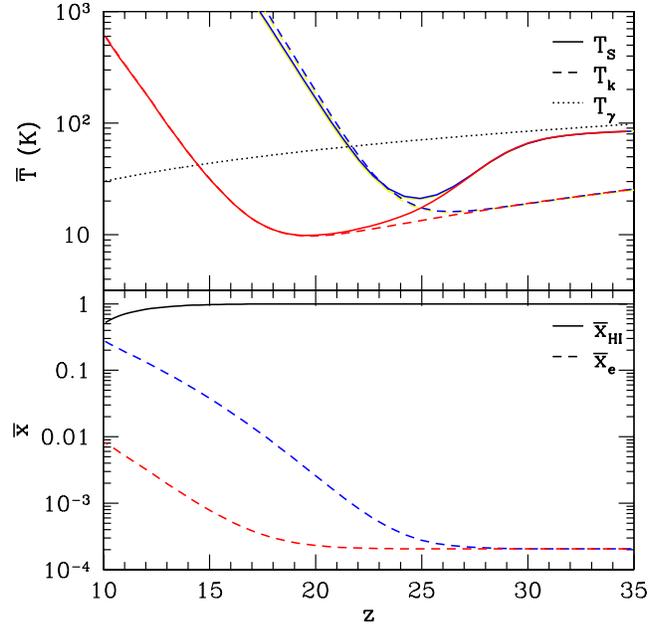}{3.3}{0.5}{.}{0.}
\caption{
{\it Top panel}: Evolution of the mean temperatures from 21cmFAST in our fiducial, $\zeta_X=10^{57} \Msun^{-1}$ model ({\it blue curves}), and a model with a hundred times weaker X-ray heating, $\zeta_X=10^{55} \Msun^{-1}$ ({\it red curves}). Solid, dashed and dotted curves show $\Ts$, $\Tk$ and $\Tcmb$, respectively.
{\it Bottom panel}: The corresponding evolution in $\bar{x}_e$ and $\avenf$. 
\label{fig:global_evolution}}
\vspace{-1\baselineskip}
\end{figure}

Thorough investigation of the available parameter space is beyond the scope of this work.  However, just to briefly show an alternate evolution, in Fig. \ref{fig:global_evolution}, we include a model where the X-ray efficiency is two orders of magnitude lower than in our fiducial model.  As one would expect, the \lya\ pumping epoch is unaffected.  However, X-ray heating is delayed by $\Delta z \sim$ 7.  In such an extreme model, the 21-cm signal would be seen in strong absorption against the CMB for a long time, and the X-ray heating epoch would overlap with the early stages of the reionization epoch.

\begin{figure*}
{
\begin{center}
\includegraphics[width=0.65\textwidth]{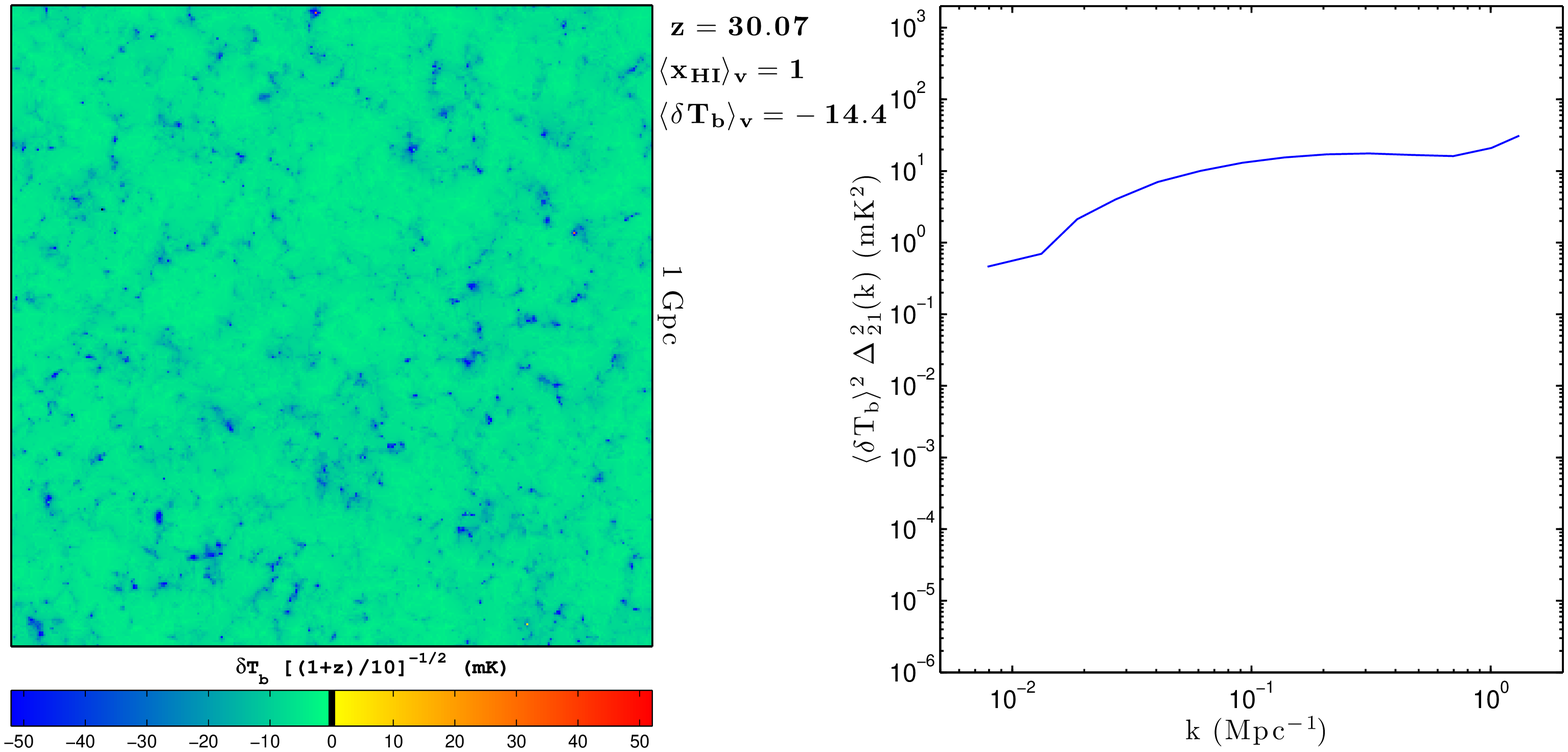}
\includegraphics[width=0.65\textwidth]{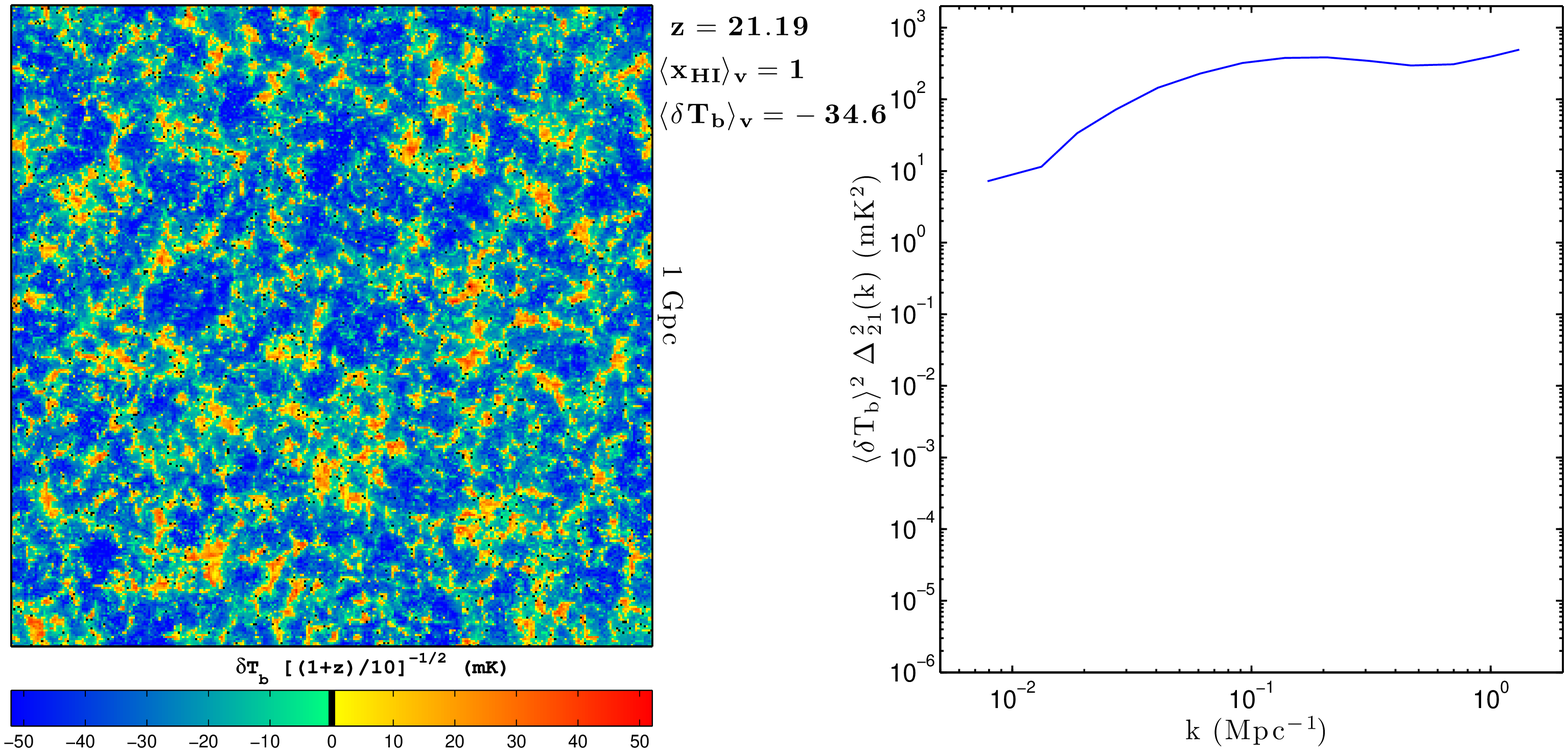}
\includegraphics[width=0.65\textwidth]{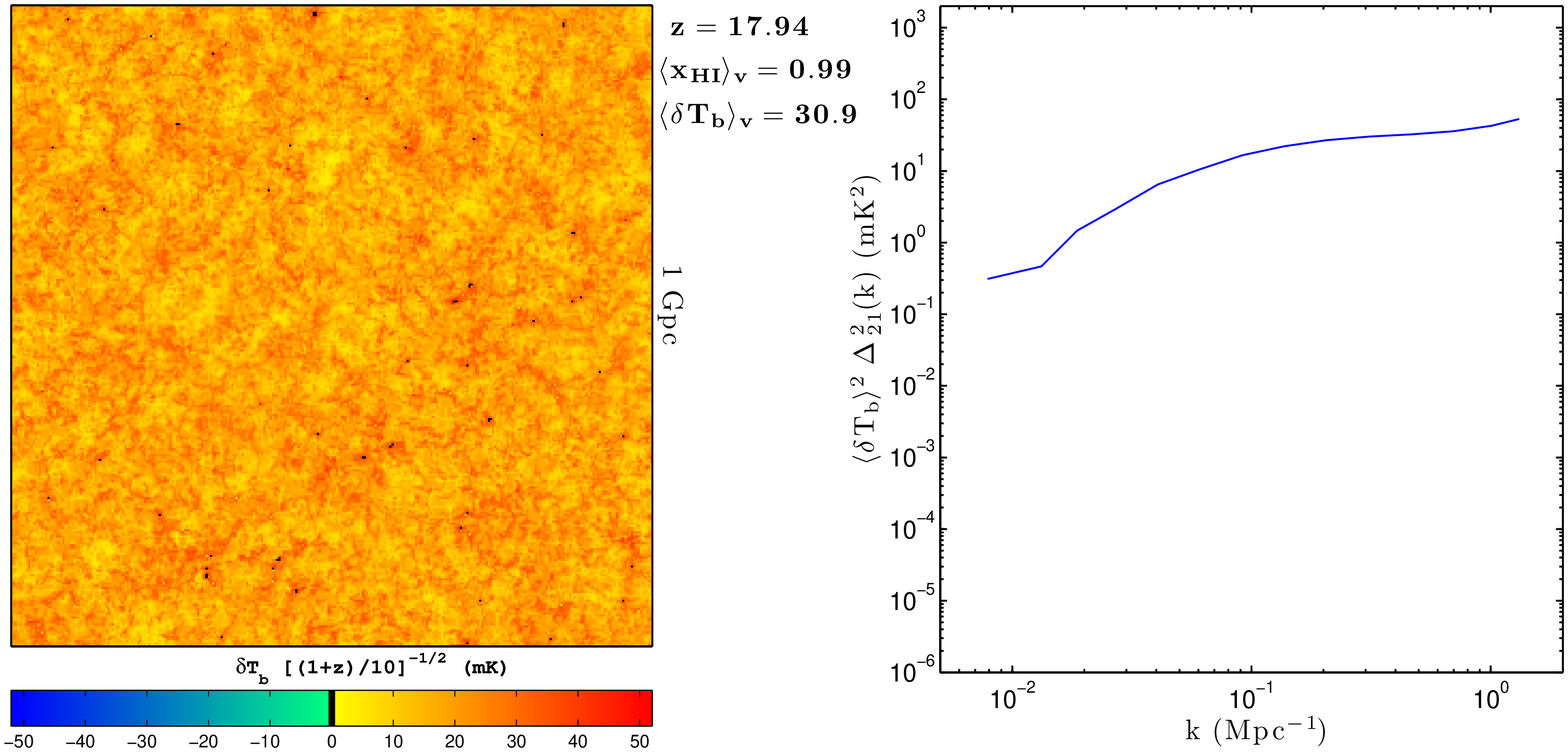}
\includegraphics[width=0.65\textwidth]{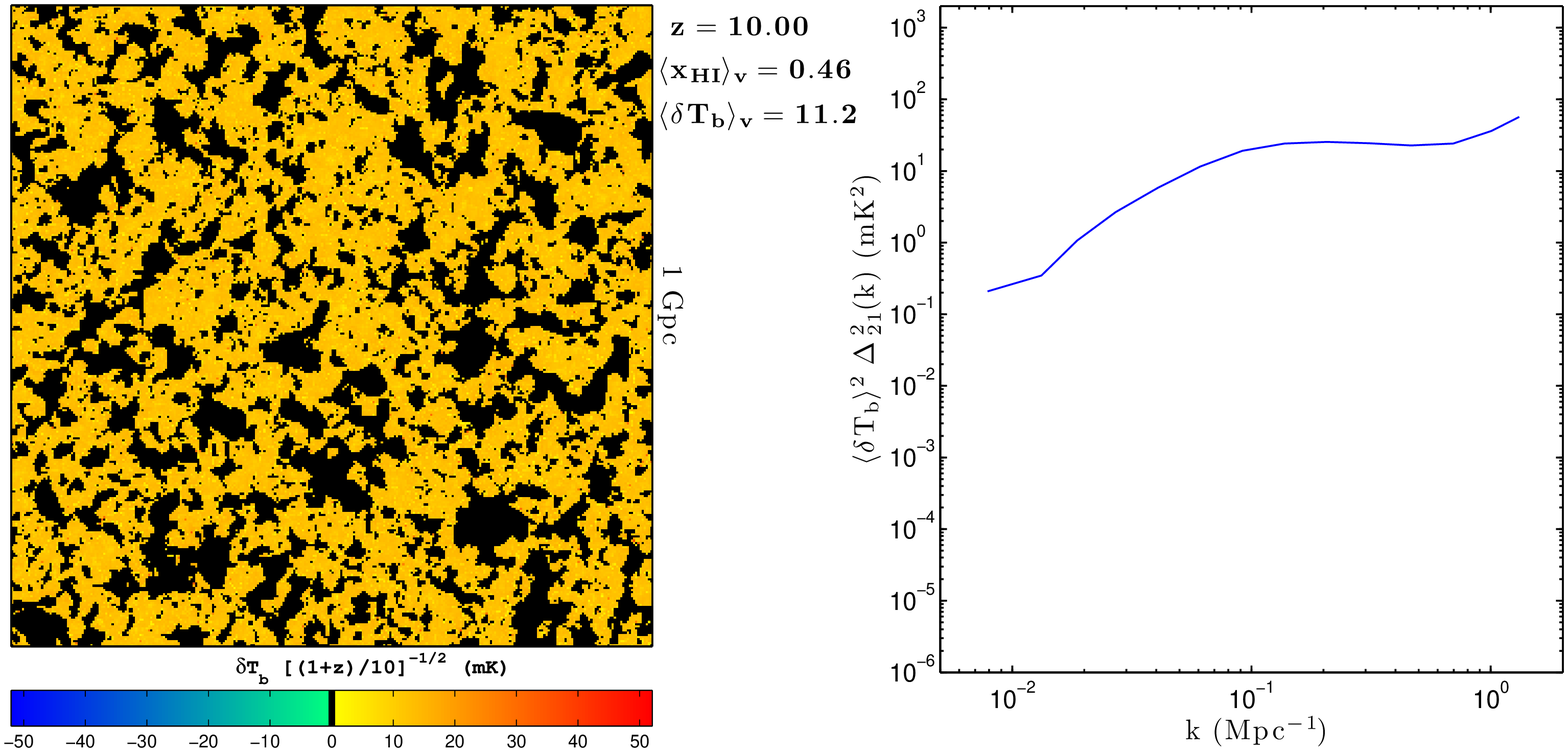}
\vskip0.0pt
\end{center}
}
\caption{
Slices through our $\delT$ simulation box ({\it left}), and the corresponding 3D power spectra ({\it right}), for our fiducial model at $z = $ 30.1, 21.2, 17.9, 10.0 ({\it top to bottom}).  The slices were chosen to highlight various epochs in the cosmic 21-cm signal (c.f. the corresponding mean evolution in Fig. \ref{fig:early_global_evolution}):  the onset of \lya\ pumping, the onset of X-ray heating, the completion of X-ray heating, and the mid-point of reionization are shown from top to bottom.  All slices are 1 Gpc on a side and 3.3 Mpc deep.  For a movie of this simulation, see http://www.astro.princeton.edu/$\sim$mesinger/Movies/delT.mov.
\label{fig:slice_withps}
}
\vspace{-1\baselineskip}
\end{figure*}

Finally, in Fig. \ref{fig:slice_withps}, we show slices through our fiducial $\delT$ box ({\it left}), and the corresponding 3D power spectra ({\it right}). The slices were chosen to highlight various epochs in cosmic 21-cm signal discussed above: the onset of \lya\ pumping, the onset of X-ray heating, the completion of X-ray heating, and the mid-point of reionization are shown from top to bottom.  We encourage the interested reader to see more evolutionary stages through the movie at http://www.astro.princeton.edu/$\sim$mesinger/Movies/delT.mov.  When normalized to the same epoch, our power-spectrum evolution agrees fairly well with the analytical model of \citet{PF07}, as well as its application to a numerical \citep{Santos08} and a semi-numerical (MF07) simulation (\citealt{Santos09}; c.f. see their Fig. 11).

One interesting feature worth mentioning is that our reionization power spectra show a drop in power on large scales, which persists throughout reionization.  In the advanced stages of reionization ($\avenf\lsim 0.9$), the strongest imprint on the 21-cm power spectrum is from HII morphology, with a ``shoulder'' feature quickly propagating from small to large scales, and flattening the power spectrum (e.g. \citealt{McQuinn07}; see Fig. \ref{fig:compare_ps}).  However there should still be a drop in large-scale power beyond either the HII bubble scale, or the photon mean free path in the ionized IGM\footnote{Note that RT simulations of reionization generally do not explicitly include an effective ionizing photon mean free path from unresolved LLSs.}, whichever is smallest. This is an interesting feature from which one can deduce the ionizing photon mean free path in the late stages of reionization \citep{MD08, FM09}.  And since it occurs at $k\lsim0.1$ Mpc$^{-1}$ (for our fiducial choice of parameters), it is beyond the dynamic range of present-day numerical simulations hoping to resolve atomically-cooled source halos.

\section{Conclusions}
\label{sec:conc}

We introduce a powerful new semi-numeric modeling tool, {\it 21cmFAST}, designed to efficiently simulate the cosmological 21-cm signal. 
Our approach uses perturbation theory, excursion set formalism, and analytic prescriptions to generate evolved density, ionization, peculiar velocity, and spin temperature fields, which it then combines to compute the 21-cm brightness temperature. This code is based on the semi-numerical simulation, DexM, (MF07).  However, here we bypasses the halo finder and operate directly on the evolved density field, thereby increasing the speed and decreasing memory requirements. 
In the post-heating regime, 21cmFAST can generate a realization in a few minutes on a single processor, compared to many days on $>$ 1000-node supercomputing cluster required to generate the same resolution boxes using state-of-the-art numerical simulations.  21cmFAST realizations in the pre-heating regime require $\sim$ one day of computation time.  Conversely, RT simulations of the pre-heating regime which resolve most sources currently do not exist, as they are too computationally expensive.
Our code is publicly available at http://www.astro.princeton.edu/$\sim$mesinger/Sim.html

We compare maps, PDFs and power spectra from 21cmFAST, with corresponding ones from the hydrodynamic numerical simulations of \citet{TCL08}, generated from the same initial conditions. 
 We find good agreement with the numerical simulation on scales pertinent to the upcoming observations ($\gsim$ 1 Mpc).  The power spectra from 21cmFAST agree with those generated from the numerical simulation to within 10s of percent down to the Nyquist frequency.

We find evidence that non-linear peculiar velocity effects enhance the 21-cm power spectrum, beyond the expected geometric, linear value.  This enhancement quickly diminishes during the onset of reionization, remaining only on small scales at $\avenf \lsim 0.7$.  Interestingly, we also find that the large-scale power is {\it decreased} as a result of peculiar velocities in the advanced stages of reionization.  The reason for this is due to the ``inside-out'' nature of reionization on large-scales: the remaining HI regions are preferentially underdense, in which peculiar velocities decrease the 21-cm optical depth and brightness temperature.

Our code can also simulate the pre-reionization regime, including the astrophysical processes of X-ray heating and the WF effect.  We show results from a 1 Gpc simulation which tracks the cosmic 21-cm signal down from $z=250$, highlighting the various interesting epochs.

There are several large 21-cm interferometers scheduled to become operational soon.  Interpreting their upcoming data will be difficult since we know very little about the astrophysical processes at high redshifts.  Furthermore, there is an enormous range of scales involved, making numerical simulations too slow for efficient parameter exploration.  21cmFAST is not.

\vskip+0.5in

We thank Hy Trac for permission to use the initial conditions and the output from his radiative transfer hydrodynamic simulation.  We thank Dave Spiegel for his considerable Matlab experience, without which making 21-cm movies would have been considerably more difficult.
 Support for this work was provided by NASA through Hubble Fellowship grant HST-HF-51245.01-A to AM, awarded by the Space Telescope Science Institute, which is operated by the Association of Universities for Research in Astronomy, Inc., for NASA, under contract NAS 5-26555.
SRF was partially supported by the David and Lucile Packard Foundation and by NASA through the LUNAR program. The LUNAR consortium (http://lunar.colorado.edu), headquartered at the University of Colorado, is funded by the NASA Lunar Science Institute (via Cooperative Agreement NNA09DB30A) to investigate concepts for astrophysical observatories on the Moon.
Support was also partially provided by
NASA grants NNG06GI09G and NNX08AH31G to RC.

\bibliographystyle{mn2e}
\bibliography{ms}

\end{document}

%% file: defns.tex
\newcommand{\Ndot}{\dot{N}_{\rm X}}
\newcommand{\Omb}{\Omega_{\rm b}}
\newcommand{\xz}{({\bf x}, z)}
\newcommand{\xzp}{({\bf x}, z')}
\newcommand{\Ts}{T_{\rm S}}
\newcommand{\Tk}{T_{\rm K}}
\newcommand{\aveTs}{\bar{T}_{\rm S}}
\newcommand{\aveTk}{\bar{T}_{\rm K}}
\newcommand{\denps}{\Delta^2_{\delta \delta}({\bf k}, z)}

\newcommand{\nf}{x_{\rm HI}}
\newcommand{\avenf}{\bar{x}_{\rm HI}}

\newcommand{\lya}{Ly$\alpha$}

\newcommand{\Msun}{M_\odot}
\newcommand{\Tvir}{T_{\rm vir}}
\newcommand{\fcoll}{f_{\rm coll}({\bf x}, z, R)}

\newcommand{\Tcmb}{T_\gamma}
\newcommand{\delT}{\delta T_b}

\newcommand{\delNL}{\delta_{\rm nl}}
\newcommand{\Mmin}{M_{\rm min}}

\newenvironment{packed_enum}{
\begin{enumerate}
  \setlength{\itemsep}{1pt}
  \setlength{\parskip}{0pt}
  \setlength{\parsep}{0pt}
}{\end{enumerate}}